\documentclass{article}
\usepackage[a4paper, total={6in, 8in}]{geometry}
\usepackage{graphicx} 
\usepackage{amsmath}
\usepackage{amsfonts}
\usepackage{color}
\usepackage{ulem}

\newcommand{\un}{{\mathbb I}}
\newcommand{\ra}{\rightarrow}
\newcommand{\tr}{{\rm tr}}

\newcommand{\ran}{\rm Ran\, }
\renewcommand{\ker}{{\rm Ker\, }}
\newcommand{\bra}{\langle}
\newcommand{\ket}{\rangle}
\newcommand{\be}{\begin{equation}}
\newcommand{\ee}{\end{equation}}
\newcommand{\bea}{\begin{eqnarray}}
\newcommand{\eea}{\end{eqnarray}}

\newcommand{\e}{{\rm e}}

\newcommand{\spec}{{\rm spec}}

\newcommand{\ffi}{\varphi}
\newcommand{\sign}{\mbox{sign}}
\newcommand{\ep}{\hfill  {\vrule height 10pt width 8pt depth 0pt}}

\newcommand{\grintl}{[\kern-.18em [}
\newcommand{\grintr}{]\kern-.18em ]}

\newcounter{resultcounter}[section]

\newtheorem{thm}[resultcounter]{Theorem}
\newtheorem{lem}[resultcounter]{Lemma}
\newtheorem{prop}[resultcounter]{Proposition}

\newtheorem{definition}[resultcounter]{Definition}
\newtheorem{rem}[resultcounter]{Remark}
\numberwithin{equation}{section}
\def\bed{\begin{definition}}
\def\eed{\end{definition}}
\def\one{{\mathchoice {\rm 1\mskip-4mu l} {\rm 1\mskip-4mu l} {\rm 1\mskip-4.5mu l} {\rm 1\mskip-5mu l}}}
\def\proof{\noindent{\bf Proof}\ }
 \def\cB{{\cal B}} \def\cC{{\cal C}}
\def\cD{{\cal D}}  
 \def\cH{{\cal H}} \def\cI{{\cal I}}
\def\cJ{{\cal J}}  \def\cL{{\cal L}}
\def\cM{{\cal M}}  
  \def\cR{{\cal R}}
 \def\cT{{\cal T}}

\newcommand{\R}{{\mathbb R}}

\newcommand{\C}{{\mathbb C}}

\newcommand{\qrm}{QRM }
\newcommand{\qrms}{QRMs}
\newcommand{\diag}{{\rm Diag}}
\newcommand{\offdiag}{{\rm Offdiag}}

\newcommand{\rank}{{\rm Rank\,}}

\begin{document}

\title{Entropy Production of Quantum Reset Models}
\author{G\'eraldine Haack\footnote{Department of Applied Physics, University of Geneva, 1211 Geneva, Switzerland} \ $\&$ Alain Joye\footnote{ Univ. Grenoble Alpes, CNRS, Institut Fourier, F-38000 Grenoble, France} }
\date{\today}
\maketitle

\begin{abstract} 

We analyze the entropy production of Quantum Reset Models (\qrms) corresponding to quantum dynamical semigroups driven by Lindbladians motivated by a probabilistic description of dissipation in an external environment. We investigate the strict positivity of entropy production for Lindbladians given as sums of \qrms, when the Hamiltonian of the total Lindbladian is split as an affine combination of Hamiltonians of the individual \qrms. In this setup, we derive conditions on the coefficients of the combination and on the reset states ensuring either positive or zero entropy production. Second, we deal with a tri-partite system subject at its ends to two independent \qrms\ and a weak coupling Hamiltonian. The latter is split as an affine combination of individual Hamiltonians, and we provide necessary and sufficient conditions ensuring strict positivity of the entropy production to leading order, with the possible exception of one affine combination. We apply these results to a physically motivated model and exhibit explicit expressions for the leading orders steady-state solution, entropy production and entropy fluxes. Moreover, these approximations are numerically shown to hold beyond the expected regimes.

\end{abstract}

\section{Introduction}

Apart from rare exceptions or specific regimes, the Hamiltonian dynamics of quantum systems interacting with a large quantum environment is out of reach, analytically and numerically. Therefore one resorts to various approximations to get insights on such dynamical systems, depending on the type of information one wishes to get. A popular approach used when the focus is on the quantum system and not on the environment, consists in adopting an approximate dynamics for the sole quantum system of interest, taking the environment into account in an effective way, through certain of its characteristics only. Indeed, tracing out the degrees of freedom of the environment, and adopting the Born-Markov approximation, leads to the celebrated Lindblad evolution equation for the quantum system of interest, \cite{L, GKS}. This linear equation defines a completely positive trace preserving (CPTP) semigroup on density matrices, the set of non-negative trace class operators of trace one. This approach is quite successful, see {\it e.g.} \cite{Breuer, AJP}, and it can be made rigorous in many circumstances, \cite{DerFru, Merkli20} for example. Another successful approach of this general question is based on repeated interaction schemes or collisional models, see for example \cite{Barra15, Barra17, Lorenzo17, Pezzuto16, Rau63,  Strasberg17, Seah19} and
\cite{AttalPautrat, BJM0, BJMRev, BruPil, HJPR,  BJP} for rigorous analyses of such models.

The generator of the Lindblad equation, called the Lindbladian, consists in a dissipator which encodes the effect of the environment, and a Hamiltonian part that describes the dynamics of the system of interest in absence of coupling to the environment. Therefore, the detailed design of the dissipator depends on the properties of the environment one considers, and there is a host of different dissipators in the literature, of a more or less sophisticated structure, that are used for various purposes. While the adoption of the Lindblad evolution equation to approximate the quantum dynamics of the full system is a huge simplification, its resolution is by no means trivial, in particular since its generator acting on the set of trace class operators lacks symmetries, {\it e.g. } self-adjointness. 

Among the possible choices of Lindbladians, the so called Quantum Reset Models (\qrms) whose dissipators are parameterized by a reset density matrix and a positive reset rate are both relevant and simple enough to allow for a rigorous treatment. In a previous work, Ref.~\cite{HJ}, the authors have performed a perturbative analysis of \qrms\ in the particularly relevant setup of multi-partite structures, driven at their ends by \qrms. Such models are used for instance to describe spin-boson chains  with nearest-neighbour interactions, amenable to experiments in solid-state physics, see {\it e.g.} \cite{Wiel02, Rubinsztein23}. \qrms \ have also proven to be particularly useful to demonstrate how thermal resources can be exploited for quantum information tasks in the context of quantum thermodynamics \cite{Skrzypczyk11, Brask15, Tavakoli18, Tavakoli20, Kulkarni23}. In Ref.~\cite{HJ}, we showed uniqueness and analyticity in a perturbative parameter of the steady state of tri-partite \qrms \ and we provided the first terms of its expansion.

\medskip

In the present paper, we explore the entropy production of various \qrms, partitioned or not.
We consider Lindbladians given as a sum of individual \qrms, each of which coming with its own stationary state. Adopting the general definition of entropy production of a state for Lindblad dynamics given in the seminal works of Lebowitz and Spohn, and Spohn, \cite{LS, S}, the entropy production of the stationary state of the full Lindbladian is then given by the sum of the entropy productions of the individual \qrms. The strict positivity of the entropy production of the stationary state is an indicator of its non-equilibrium nature, and it is the primary objective of this work. See the paper \cite{JPW} for an general overview of entropic considerations for quantum dynamical semigroups. 

We address this question in two scenarios, providing explicit descriptions of the asymptotic states of the individual Lindbladians and of the total Lindbladian. First, we suppose the Hamiltonian of the total Lindladian is given by an affine combination of the Hamiltonians of the individual \qrms \ that we assume are all proportional, with otherwise unrelated reset states and rates. We provide conditions on the coefficients and reset states ensuring either positive or zero entropy production of the stationary state, and we discuss their links with detailed balance conditions. Second, we consider a structured tri-partite Hilbert space and two \qrms \ with dissipators acting on different and independent parts of the Hilbert space as addressed in \cite{HJ}. The total Hamiltonian is again an affine combination of the individual Hamiltonians assumed to be proportional, and it weakly couples all three parts of the Hilbert space. 
 The strength of the coupling is monitored by a small coupling constant in front of the Hamiltonian, while the dissipators are of order one. In this regime, the asymptotic state of such structured \qrms\ is shown in \cite{HJ} to be unique and analytic in the coupling constant. For weak enough coupling, we prove that the leading order entropy production, proportional to the square of the coupling constant, is strictly positive (with the possible exception of one affine combination) if and only if the commutator of the Hamiltonian with the leading term of the stationary state is non zero. 
 Our results are applied to a physical model, for which we make explicit quantities and concepts introduced in the theorems, lemma and proofs. Interestingly, the model allows us to go beyond certain general statements, for instance demonstrating the validity of leading order expressions to higher orders than mathematically proven 
for the entropy production and entropy fluxes. Exact expressions also allow us to derive equilibrium-necessary conditions for zero entropy production, as well as conditions under which the sign of the entropy fluxes may change on this model.

The paper is organized as follows. The next section recalls the general notions of entropy production of states for Lindbladians given by sums of individual Lindbladians. Section \ref{sec:qrm} introduces Quantum Reset Models and discusses the detailed balance condition for them, as well as some properties of their asymptotic states. The positivity properties of the entropy production for certain affine combinations of Hamiltonians are presented in Section \ref{sec:affine}, while Section \ref{exandnum} is devoted to examples and further numerical investigations.  
The setup for tri-partite systems is recalled in Section \ref{sec:struct}, while the perturbative analysis of the entropy production in this framework is presented in Section \ref{sec:smalldrive}. The analysis of a realistic three-qubit model, including numerical investigations, closes the paper.

\section{Entropy production in the Lindbladian setup}\label{genlindframe}
We recall here the concepts of Entropy Production  for Quantum Dynamical Semigroups generated by a sum of Lindblad operators.
\medskip

The effective evolution equation of a state within Lindblad's framework reads 
\begin{align}\label{lindblad}
\dot{\rho}(t)=\cL (\rho(t)), \ \ t\in (0,\infty), \ \ \rho(0)=\rho_0\in \cT(\cH),
\end{align}
where 
\begin{align}\label{genlib}
\cL(\cdot)=-i[H,\cdot ]+\sum_{l\in \cI} \Big(\Gamma_l\cdot \Gamma_l^*-\frac12\big\{\Gamma_l^*\Gamma_l, \cdot\big\}\Big).
\end{align}
Here $H$ and $\Gamma_l$, $l\in \cI$ are all bounded operators on the Hilbert space $\cH$ and $H$ is self-adjoint, $\cT(\cH)$ denotes the set of trace class operators and $\rho(t)$ is the state of the system at time $t\in[0,\infty)$. The Hamiltonian part of the Lindbladian (\ref{genlib}), given by the commutator with $H$, describes the evolution of the state in absence of environment, while the global effect of the latter on the evolution is encoded in the dissipator, given by the second term in (\ref{genlib}). Moreover, the map $\e^{t\cL}$ on $\cT(\cH)$ is trace preserving and completely positive, which defines a Quantum Dynamical Semigroup (QDS) $(e^{t\cL})_{t\geq 0}$.

The set of states is denoted by $\cD\cM(\cH)=\{\rho\in \cB(\cH)\, |\, \rho=\rho^*\geq 0, \, \tr(\rho)=1\}$. Under quite general circumstances, if $\rho_0\in \cD\cM(\cH)$, the state at time $t>0$, $\rho(t)=\e^{t\cL}(\rho_0)$, converges exponentially fast as $t\ra\infty$ to a unique faithful ({\it i.e.} strictly positive) steady state denoted by $\rho^+\in \cD\cM(\cH)$ that belongs to the kernel of $\cL$. 

\medskip

We are interested in the situation where the Lindbladian $\cL$ admits a decomposition 
\be\label{decomplind}
\cL=\sum_{j\in \cJ} \cL_j, 
\ee
where each $\cL_j$ is a Lindblad operator possessing a unique faithful steady state $\rho_j^+$. The interpretation is that each individual Lindbladian $\cL_j$ describes the interaction of the system with a reservoir $\cR_j$.

Following Lebowitz and Spohn, see \cite{LS, JPW},  we consider the entropy production (EP) of the QDS $(e^{t\cL})_{t\geq 0}$  in the state $\rho$ defined by
\be\label{entprodgen}
\sigma(\rho)=-\frac{d}{dt}\sum_{j\in \cJ}S(e^{t\cL_j}(\rho)|\rho_j^+)|_{t=0},
\ee
where $S(\mu |\nu)$ denotes the relative entropy of the state $\mu$ with respect to the state $\nu$ given by
\be\label{relent}
S(\mu |\nu)=\left\{ \begin{matrix} \tr (\mu (\ln(\mu)-\ln(\nu) ) & {\rm if} \ \ker (\nu)\subset \ker (\mu) \\
+\infty & {\rm otherwise.}\end{matrix}  \right.
\ee

As proven in  \cite{LS, S}, the EP enjoys the following properties:
\begin{align}
\sigma(\rho)\geq 0, 
\end{align}
and as a map on $\cal DM(H)$, the set of states or density matrices on $\cH$, $\rho\mapsto \sigma(\rho)$ is convex. 

Moreover, with $S(\rho)$ the entropy of the state $\rho\in \cal DM(H)$,
\be\label{ent}
S(\rho)=-\tr (\rho\ln(\rho))\geq 0,
\ee 
we have the relation
\be\label{entbal}
\frac{d}{dt}S(e^{t\cL}(\rho))|_{t=0} = \sigma(\rho)-\sum_{j\in\cJ}\tr(\cL_j(\rho)\ln(\rho_j^+))\,,
\ee
also called entropy balance equation.
With $\cL^\dagger_j$, the adjoint of $\cL_j$ with respect to the Hilbert-Schmidt scalar product of matrices, the summands in the last term of the RHS can be written as 
\be\label{entcur}
\tr (\rho \cL^\dagger_j(-\ln(\rho_j^+)))\equiv \tr (\rho \cI_j^+),
\ee
where the observable $\cI_j^+=\cL^\dagger_j(-\ln(\rho_j^+))$ is interpreted as the entropy flux out of the $j^{\rm th}$ reservoir.

Finally, the following expression for the EP will be the starting point of our analysis below
\be\label{entprodfor}
\sigma(\rho)=\sum_{j\in\cJ}\tr(\cL_j(\rho)(\ln(\rho_j^+)-\ln(\rho)))\equiv \sum_{j\in\cJ} \sigma_j(\rho),
\ee
where each $ \sigma_j(\rho)\geq 0$ is the EP of the state $\rho$ for the individual QDS $(e^{t\cL_j})_{t\geq 0}$.

Accordingly, in case there is a unique reservoir and $\cL$ admits $\rho^+\in {\cal DM(H)}$ as its unique steady state, the EP of of the state $\rho$ for the QDS $(\e^{t\cL})_{t\geq 0}$ is denoted by $\sigma_\cL(\cdot)$ and reads
\begin{align}\label{EPL}
    \sigma_\cL(\rho)=\tr(\cL(\rho)(\ln(\rho^+)-\ln(\rho))).
\end{align}

Assuming that  $\cL$ admits $\rho^+\in {\cal DM(H)}$ as its unique steady state, the analysis in the Lindbladian framework of the multireservoir case focuses on the strict positivity  
of the EP in the steady state which reads
\begin{align}
\label{entsteady}
\sigma(\rho^+)&=\sum_{j\in\cJ}\tr(\cL_j(\rho^+)(\ln(\rho_j^+)-\ln(\rho^+)))\nonumber \\
&=\sum_{j\in\cJ}\tr(\cL_j(\rho^+)\ln(\rho_j^+)),
\end{align}
thanks to (\ref{decomplind}) applied to $\rho^+$. Note, however, that the individual summands 
\begin{align}
\tr(\cL_j(\rho^+)\ln(\rho_j^+))=-\tr(\rho^+ \cI_j^+)
\end{align}
corresponding to the entropy fluxes into the $j^{\rm th}$ reservoir, are {\it not} necessarily positive.
Finally, in case there is only one reservoir, the EP of the asymptotic state $\rho^+$ vanishes,  $\sigma_\cL(\rho^+)=0$,  as expected. 

\section{Quantum Reset Models on a Simple Hilbert Space}\label{sec:qrm}
In this section we address the entropy production for a Lindblad operator defined as a sum of Quantum Reset Models (QRM), all defined on a Hilbert space $\mathcal{H}$ with no particular structure. By contrast, we consider in Section \ref{sec:struct} below the richer situation of a tri-partite system defined on the tensor product of three Hilbert spaces, in the spirit of \cite{HJ}.\\

Let us consider a single quantum system of finite dimension, with Hamiltonian $H$ defined on a Hilbert space $\cH$ that is coupled to $J$ reservoirs. 
The QRM evolution equation for the state is defined by:
\begin{align}\label{simpleqrm}
\dot{\rho}(t)=-i [H, \rho(t)]+\sum_{j\in \mathcal{J}}\gamma_j(\tau_j \, \tr(\rho(t))-\rho(t))\, , 
\end{align}
where $\mathcal{J}=\{1,2\dots, J\}.$
The operator $\rho\in \mathcal{DM(H)}$ is the state of the system defined on $\cH$, $\gamma_j\geq 0$ is the coupling rate to the reservoir $j\in \mathcal{J}$, whose  asymptotic state is $\tau_j\in{\cal DM(H)}$, in absence of Hamiltonian. 
\medskip

We assume the following hypotheses on the dissipator of the generator (\ref{simpleqrm}) in the present section.

\noindent
{\bf Diss}:\\
Let   $\cH$ be a Hilbert space, with $\dim \cH=d<\infty$. The dissipator is characterised by  
\begin{itemize}
\item $\{\tau_j\}_{j\in\mathcal{J}}$ a collection of density matrices on $\cH$, {\it i.e.} $\tau_j\in \cD\cM(\cH)$, for all $j\in \mathcal{J}$, 
\item $\gamma_j >0$,  $j\in \mathcal{J}$, the collection of associated non{-}zero rates for the coupling to the $J$ reservoirs.
\end{itemize}
The Hamiltonian part of the generator, $H=H^*\in\cB(\cH)$, is arbitrary so far, and we denote its spectrum by $\sigma(H)=\{e_1, e_2, \dots, e_d\}$, with eigenvalues repeated according to their multiplicity.

\medskip

The generator of the \qrm is the Lindbladian $\cL\in\cB(\cB(\cH))$ defined by 
\begin{align}\label{simpleqrmgen}
\cL (\rho)=-i [H, \rho]+\sum_{j\in \mathcal{J}}\gamma_j(\tau_j \, \tr(\rho)-\rho)
\end{align}
where $\rho$ here is arbitrary in $\cB(\cH)$, such that the dynamics of the \qrm reads
\begin{align}\label{dynqrm}
\dot{\rho}(t)=\cL (\rho(t)), \ \ t\in (0,\infty), \ \ \rho(0)=\rho_0\in \cB(\cH).
\end{align}
In case $\rho\in {\cal DM(H)}$ the trace factor in (\ref{simpleqrmgen}) equals one.
The generator $\cL$ is well known to have the form of a Lindbladian dissipator, see also (\ref{ldagger}) below.

\medskip

Let us quickly recall the properties of the dynamics generated by  (\ref{simpleqrm}) (see \cite{HJ} for details).
Combining the density matrices $\tau_j$ with corresponding rates $\gamma_j$ into a single density matrix $T$ with corresponding rate $\Gamma$ according to
\be\label{recomb}
\Gamma=\sum_{j\in \mathcal{J}}\gamma_j>0,  \ \ T=\frac{1}{\Gamma }\sum_{j\in \mathcal{J}} \gamma_j \tau_j\in {\cal DM(H)},
\ee
(\ref{simpleqrmgen}) simplifies to
 \be\label{renl}
 \cL(\rho)=-i[H,\rho]+\Gamma (T\tr (\rho)-\rho).
 \ee
 The \qrm  Lindbladian, as a linear operator acting on $\mathcal{B(H)}$ has spectrum
 \be\label{speclind}
 \spec (\cL)=\{0,-\Gamma\}\cup\{-\Gamma-i(e_j-e_k)\}_{1\leq j\neq k\leq d},
 \ee
 where $0$ is a simple eigenvalue, $-\Gamma$ is at least $d-1$ times degenerate, and (potential) eigenvalues with non zero imaginary parts appear as complex conjugate pairs, degenerate or not, depending on $\sigma(H)$.
 Then, the solution to (\ref{simpleqrm}) reads 
 \begin{align}\label{basisindep}
\rho(t)=&\e^{-t(i[H,\cdot]+\Gamma)}\big(\rho_0- \tr (\rho_0)\Gamma \big(i[H,\cdot]+\Gamma\big)^{-1}(T) \big)\nonumber\\
&+\tr (\rho_0)\Gamma \big(i[H,\cdot]+\Gamma\big)^{-1}(T),
\end{align}
which, in case $\rho_0\in {\cal DM(H)}$ yields the asymptotic state
\begin{align}\label{rhodt}
\rho^{+} \equiv \lim_{t\ra \infty}\rho(t)&=\Gamma \big(i[H,\cdot]+\Gamma\big)^{-1}(T)\nonumber\\
&= \big(i[H/\Gamma,\cdot]+1\big)^{-1}(T)
\in {\cal DM(H)}.
\end{align}
In particular, $\ker \cL$ is thus spanned by $\rho^+$, and if $[T,H]=0$, then $\rho^{+} =T$. 

Denoting by $\{\ffi_l\}_{1\leq l\leq d}$ the normalised eigenvectors of $H$ associated to $\spec(H)=\{e_1, e_2, \dots, e_d\}$, 
we write $A_{mn}=\bra \ffi_m | A \ffi_n\ket$ the corresponding matrix elements of $A\in \cB(\cH)$. Hence,
 \begin{align}\label{rhoft}
 \rho^{+}_{mn}={\big(i[H/\Gamma,\cdot]+1\big)^{-1}(T)}_{mn}= \frac{T_{mn} }{(i(e_m-e_n)/\Gamma+1)}.
 \end{align}
 Introducing the spectral projectors of $H$, $P_j$ associated with the eigenvalue $e_j$, we can write
 \begin{align}\label{rpspecpro}
     \rho^+=\sum_{j,k}\frac{P_j T P_k}{(i(e_j-e_k)/\Gamma+1)}.
 \end{align}
 Observe that $\rho^+$ depends linearly on $T$ and that it depends on the ratio $H/\Gamma$.

 We now state some properties regarding the asymptotic state $\rho^+$:
 \begin{lem} \label{lem:posrho}
    For $H=H^*$, and $T\in \mathcal{B(H)}$, the linear map  $\rho_H: \cB(\cH)\ra \cB(\cH)$ $$T\mapsto \rho_H(T)= \big(i[H,\cdot]+1\big)^{-1}(T),$$ see (\ref{rhodt}), is CPTP and satisfies 
    \begin{align}
        T>0 \ \Rightarrow \  \rho_H(T)>0. 
    \end{align}
    Moreover, for $0<T\in \mathcal{DM(H)}$, 
    \begin{align}
    &\tr(T\ln(\rho_H(T)))=\tr (\rho_H(T)\ln(\rho_H(T))), \nonumber\\
        &S(\rho_H(T))=S(T)+S(T|\rho_H(T)). \label{tbg}
    \end{align}
    Finally, with $^t$ denoting transposition,
    \begin{align}\label{h-h}
        \overline{\rho_{H}(T)}={\rho_{-H}(\overline{T})}, \ \ \rho_{H}^t(T)={\rho_{-H}(T^t)}.
    \end{align}
\end{lem}
\begin{rem}
    The commutant $H'=\{ T\in \cB(\cH), \ | \ [T,H]=0\}$ coincides with  $\ker \rho_H(\cdot)-\un$, which shows that $\rho_H$ is not positivity improving. Also, $\rho_H(\cdot)$ increases the entropy strictly, unless $T\in H'$ is invariant.
\end{rem}
\proof:
The linear map $\mathcal{B(H)}\ni T\mapsto \rho_H(T)$ satisfies the identity
\begin{align}\label{laplace}
    \rho_H(T)=\int_0^\infty \e^{-t}\e^{-it[H,\cdot ]}(T) dt=\int_0^\infty \e^{-t/2}\e^{-itH}T\e^{itH}\e^{-t/2} dt,
\end{align}
where the integral is convergent in norm, and
$
\int_0^\infty \e^{-t/2}\e^{+itH}\e^{-itH}\e^{-t/2} dt=1.
$
This continuous Kraus decomposition shows that $\rho_H$ is a CPTP map, see {\it e.g.} Section 2 in \cite{RS}.

Consider now $T>0$ and its spectral decomposition of $T=\sum_{j} t_j P_j$, where $j$ labels the eigenvalues of $T$, $t_j>0$ and $P_j\geq 0$ denote the spectral projectors. Suppose $\chi\in \cH$ is a non zero vector such that $\bra \chi | \rho_H(T)\chi\ket=0$. By linearity and positivity of the map $\rho_H(\cdot)$, using  $t_j>0$, we get that $\chi$ satisfies $\bra \chi | \rho_H(P_j)\chi\ket=0$, $\forall j$. Summing over $j$ and noting that $\rho_H(\one)= \one$, we deduce that $\|\chi\|^2=0$, a contradiction.

Then, identity
\be\label{idrhoh}
\rho_H(T)+i[H,\rho_H(T)]=T,
\ee
implies by cyclicity of the trace and $[\rho_H(T),\ln (\rho_H(T))]=0$ that
\be\label{useful}
\tr (T\ln(\rho_H(T)))=\tr(\rho_H(T)\ln(\rho_H(T))).
\ee
Thus,
\begin{align}
    S(T|\rho_H(T))=-S(T)-\tr (T\ln(\rho_H(T))=-S(T)+S(\rho_H(T)).
\end{align}
The last statement follows from the explicit representation (\ref{rhoft}).
 \ep
\medskip

Finally, we note that $\cL^\dagger: \cB(\cH)\ra \cB(\cH)$, the adjoint of $\cL$ with respect to the Hilbert Schmidt scalar product on $\cB(\cH)$ that generates the dynamics of observables, acts on $X\in \cB(\cH)$ as
\begin{align}\label{ldagger}
    \cL^\dagger (X)&=i[H,X]+\Gamma (\un \, \tr (T X)-X)\nonumber\\
    &=i[H,X]-\frac{1}{2}\{\Phi(\un), X\}+\Phi(X),
\end{align}
where $\Phi: \cB(\cH)\ra \cB(\cH)$ is the CP map defined by 
\begin{align}\label{defficp}
    \Phi(X)=\Gamma \un \, \tr (TX).
\end{align}
The structure (\ref{ldagger}) ensures $\cL$ has Lindblad form and is convenient to assess conditions under which the pair $(\rho^+,\cL)$ satisfies the detailed balance condition, see Definition 2.4 in \cite{JPW}, which we repeat below. The detailed balance condition, ({\bf DB}), characterizes the equilibrium between the system and its environment. In our case, the former is described by the Hamiltonian $H$, while the latter is encapsulated in the dissipator $\Gamma (T\tr (\cdot) -\cdot )$.

Let $0<\rho\in \mathcal{DM(H)}$, and consider the $\rho$ dependent scalar product on $\mathcal{B(H)}$
\begin{align}\label{rhosp}
    (A,B)_\rho=\tr (\rho A^*B).
\end{align}
The adjoint of a map $\Psi\in \cB(\cB(\cH))$ with respect to this scalar product is called the $\rho$-adjoint of $\Psi$ and is denoted by $\Psi^{\rho}$.
\medskip

The definition of detailed balance reads as follows.

\medskip

\noindent ({\bf DB}): For $\cL^{\dagger}$ given by (\ref{ldagger}) and $0<\rho^+\in \ker \cL$, the detailed balance condition holds if $\Phi$ is self-adjoint with respect to the scalar product (\ref{rhosp}), {\it i.e.} if $\Phi=\Phi^{\rho^+}$. 
\medskip

If $T>0$, then $\rho^+>0$ so (\ref{rhosp}) for $\rho^+$ is a scalar product. We can characterize the conditions for detailed balance for $\cL$ given by (\ref{simpleqrmgen}), and compute its EP in case ({\bf DB}) holds: 
\begin{lem}\label{lem:DB}
    For $T>0$, the pair $(\rho^+, \cL)$ satisfies the detailed balance condition {\rm ({\bf DB})} if and only if $[T,H]=0$.\\
    If {\rm ({\bf DB})} holds, the EP of $\cL$ alone, (\ref{EPL}), reads  
\begin{align} 
\sigma_\cL(\rho)=\Gamma (S(T|\rho)+S(\rho |T)), \ \ \forall \, 0<\rho \in \mathcal{DM(H)},
\end{align}
and $\sigma_\cL(\rho)=0$ if and only if $\rho=T$.
\end{lem}
\proof:
We first check that 
$\Phi$ defined by (\ref{defficp}) satisfies for all $X\in \cB(H)$
\begin{align}
    \Phi^{\rho^+}(X)=\Gamma \, T \tr(X\rho^+)[\rho^{+}]^{-1}.
\end{align}
Then, $\Phi=\Phi^{\rho^+}$ is equivalent to $\rho^+=T$ (consider the action on $X=\un$) which, in turn, is true if and only if $[T,H]=0$, thanks to the identity $T=\rho^+ + [H,\rho^+]\frac{i}{\Gamma}$.

For the second statement, $[H,T]=0$ implies $\rho^+=T$ and we have from (\ref{EPL}) and (\ref{simpleqrmgen})
\begin{align}
    \sigma_\cL(\rho)=\tr \big( (-i[H,\rho]+\Gamma (T-\rho))(\ln(T)-\ln(\rho) \big).
\end{align}
The contribution stemming from $[H,\rho]$ vanishes thanks to the cyclicity of the trace and $[\rho,\ln(\rho)]=0=[H,T]$. Recalling (\ref{relent}), the remaining terms yield the sum of relative entropies.
\ep

\section{Specific affine combinations}\label{sec:affine}

Coming back to the entropy production in the framework of Section \ref{genlindframe}, we note that  expression (\ref{simpleqrmgen}) for $\cL$ can match the decomposition (\ref{decomplind}) $\cL=\sum_{j\in \mathcal{J}}\cL_j$ in several ways, splitting the Hamiltonian part among the different reset dissipators according to certain combinations as we now explain. This procedure admittedly entails a certain degree of arbitrariness, but it allows us to consider certain cases relevant for applications.

\medskip

Let $\{\lambda_j\}_{j\in\mathcal{J}}$, $\lambda_j\in \R$ with $\sum_{j\in\mathcal{J}}\lambda_j=1$, and $\{\tau_j\}_{j\in\mathcal{J}}$, $0<\tau_j\in \mathcal{DM(H)}$. Define
\be\label{lindj}
\cL_j(\rho)=-i [\lambda_j H, \rho]+\gamma_j(\tau_j \, \tr(\rho)-\rho),
\ee
so that (\ref{decomplind}) holds.
Stressing for now the dependence in the parameters $\lambda_j$,
\be\label{steadyj}
\rho_j^+(\lambda_j)=\gamma_j \big( i\lambda_j [H,\cdot]+\gamma_j\big)^{-1}(\tau_j)
\ee
is the unique positive steady state of the dynamics generated by $\cL_j$, with $0<\rho_j^+(\lambda_j)\in \ker \cL_j$, thanks to  Lemma \ref{lem:posrho}. 
Hence, with
\be\label{ssfd}
\rho^+=\Gamma \big(i[H,\cdot]+\Gamma\big)^{-1}(T)
\ee
the EP of the steady state of the full dynamics, $\sigma(\rho^+)$, is given by (\ref{entsteady}).
We compute 
\be\label{adjcur}
\cL_j(\rho^+)=\lambda_j \Gamma (\rho^+ -T)+\gamma_j(\tau_j-\rho^+)
\ee
so that the individual EP's read
\be\label{entprodcj}
0\leq \sigma_j(\rho^+)=\tr\Big((\lambda_j \Gamma (\rho^+ -T)+\gamma_j(\tau_j-\rho^+))\big(\ln(\rho_j^+((\lambda_j))-\ln(\rho^+)\big)\Big).
\ee
In particular, we have for $\lambda_j=0$, 
\begin{align}\label{sjlj}
    &\rho_j^+(0)=\tau_j, \ \ \cL_j(\rho^+)|_{\lambda_j=0}=\gamma_j(\tau_j-\rho^+), \nonumber\\ 
    &\sigma_j(\rho^+)|_{\lambda_j=0}=\gamma_j(S(\tau_j|\rho^+)+S(\rho^+|\tau_j))
\end{align}
where the $j$-th EP is zero if and only if $\tau_j=\rho^+$; recall the definition of the relative entropy (\ref{relent}).

\medskip

As a function of $\lambda_j$, $\sigma_j(\rho^+)$  has isolated zeros or is constant, thanks to the
\begin{lem}
   For $\tau_j>0$, the map  $\R\supset I\ni \lambda_j\ra \sigma_j(\rho^+)\in\R^+$ admits an analytic extension in a complex neighbourhood of any finite length interval  $I\subset \R$.
\end{lem}
\proof: Indeed, $\lambda_j\mapsto \rho^+_j(\lambda_j)$ is a meromorphic matrix valued function, with simple poles at $\{\frac{i\gamma_j}{e_r-e_s}\}_{1\leq r, s\leq d}$, 
 see (\ref{rhoft}), and $\rho^+_j(\lambda_j)>0$ for $\lambda_j\in I$. Hence, for any $\lambda_j\in\C$ in a small enough neighborhood of $I$, the spectrum of $\rho^+_j(\lambda_j)$ belongs to a neighborhood of $\R^+_*$, so that $\ln(\rho^+_j(\lambda_j))=-\frac{1}{2i\pi}\oint_{C}\ln(z)(\rho^+_j(\lambda_j)-z)^{-1}dz$, where $C$ is a simple positively oriented path encircing $\spec(\rho_j^+(\lambda_j))$ with $C\cap \R_-=\emptyset$, is well defined by analytic functional calculus. The analyticity of the resolvent $(\rho^+_j(\lambda_j)-z)^{-1}$  in both variables $(\lambda_j,z)$, for $z$ away from 
$\spec(\rho^+_j(\lambda_j))$ ensures analyticity of $\lambda_j\mapsto \ln(\rho^+_j(\lambda_j))$ in a neighborhood of $I\subset\R$, which yields the result.
\ep

\medskip

The non negative EP $\sigma(\rho^+)$ depends on the choice of $\{\lambda_j\}_{j\in \cJ}$ in an intricate way in general, and is likely to be zero in non generic situations only, as a sum of non negative regular functions of these parameters:
\begin{lem} The EP $\sigma(\rho^+)=0$ if and only if there exists $\{\lambda_j\}_{j\in\mathcal {J}}$, $\lambda_j\in\R$, such that  $\sigma_j(\rho^+)|_{\lambda_j}=0$ for all $j\in\mathcal{J}$, and $\sum_{j\in\mathcal{J}}\lambda_j=1$. 
\end{lem}

Yet, due to the delicate dependence of $\sigma_j(\rho^+)$ on $\lambda_j$, it is not always easy to determine the zeros of $\sigma_j(\rho^+)$.
We exhibit below combinations 
$\{\lambda_j\}_{j\in \cJ}$ which allow to assess the strict positivity of $\sigma(\rho^+)$, depending on the properties of the reset matrices $\tau_j>0$, the Hamiltonian $H$, and the parameters $\gamma_j$,  $j\in \cJ$.

\begin{prop}\label{prop:affine}
 Assume {\bf Diss} with $\tau_j>0$ $\forall j\in\mathcal{J}$. 
 Then,\\
 i) Setting $\lambda_j=\gamma_j/\Gamma\in (0,1)$ $\forall j\in\mathcal{J}$ yields 
 $\sigma(\rho^+)\geq 0$ with equality if and only if $\tau_j=\tau_k  \ \forall \, j, k\in  \mathcal{J}$. \\ 
ii) Setting $\lambda_1=1$ and $\lambda_j=0$ $\forall \, j\neq 1\in \mathcal{J}$ yields
$\sigma(\rho^+)>0$, if $\exists  \tau_j\neq \tau_k$, for $j\neq k \in \mathcal{J}\setminus \{1\}$.\\ 
iii) If $[H,\tau_j]=0$ $\forall j\in\mathcal{J}$, then for any choice of $\{\lambda_j\}$, $\rho^+=T$,  $\rho_j^+=\tau_j$ and 
$\sigma_j(\rho^+)=\gamma_j(S(\tau_j|T)+S(T|\tau_j))$. Thus $\sigma(\rho^+)\geq 0$, with equality if and only $\tau_j=T$ for all $j\in \mathcal{J}$.
\end{prop}
\begin{rem}
    Strict positivity of $\sigma(\rho^+)$ holds in a neighborhoud of the affine combinations where it is shown to hold, by continuity.
\end{rem}

We provide here some more details on these results, and some comments on their physical interpretation:

\medskip

The choice $\lambda_j=\gamma_j/\Gamma$ defining case i) is a natural convex combination, since it agrees with that defining $T$ as a function of the $\tau_j$'s and it implies that $\rho^+$ is the same convex combination of the $\rho^+_j$'s, see (\ref{comrho}).
In particular, in this case, formula 
$\sigma(\rho^+)=\sum_{j\in\mathcal{J}}\gamma_j\tr \big((\tau_j-T)\ln(\rho^+_j)\big)$
holds, see (\ref{formEP}), where each individual summand equals the entropy flux into the $j^{\rm th}$ reservoir. 
Also, further assuming $\tau_j=\tau$ for all $j\in \mathcal{J}$, we get $\cL_j=(\gamma_j/\Gamma) \cL$ so that $\rho^+_j=\rho^+$, and thus $\sigma_j(\rho^+)=0$. In a sense, all reservoirs become equivalent, which explains the zero EP. 
\\

Case ii) corresponds to attributing the Hamiltonian part to one partial Lindbladian only, $\cL_1$, while all others consist in dissipators only. This yields the expression 
$\sigma_j(\rho^+)=\gamma_j(S(\tau_j|\rho^+)+S(\rho^+|\tau_j))$ for $j\neq 1$, see (\ref{hamone}), showing that two different dissipators imply strict positivity of the total EP, and the bound
\begin{align}
    \sigma(\rho^+)\geq \sum_{j\neq 1} \gamma_j(S(\tau_j|\rho^+)+S(\rho^+|\tau_j)).
\end{align}

Thanks to Lemma \ref{lem:DB}, case iii) is equivalent to assuming condition {\bf DB} holds for all pairs $(\rho_j^+, \cL_j)$, which implies it holds also for $(\rho^+, \cL)$, and $\rho_j^+=\tau_j$, $\rho^+=T$. We actually get that  under {\bf DB}, $\forall \lambda_j\in \mathbb R$, irrespective of $\sum_{j\in \mathcal J}\lambda_j$, all individual EPs are independent of $\lambda_j$ and $\sigma_j(\rho^+)=\gamma_j(S(T|\tau_j)+S(T|\tau_j))$. Again, as soon as two reset matrices differ, $\sigma(T)>0$. In particular, if the  reset matrices are all Gibbs states at various temperatures, $\tau_j=e^{-\beta_j H}/Z_j$, {\bf DB} holds for all subsystems, and for the pair $(T, \cL)$, with $T=\sum_{j\in \mathcal J} (\gamma_j/\Gamma)e^{-\beta_j H}/Z_j$. The latter is a Gibbs state if and only if $\beta_j=\beta$ $\forall j\in \mathcal{J}$, {\it i.e.} if and only if $\sigma(\rho^+)=0$. This is in keeping with the fact that equilibrium takes place in this thermal situation if and only if the temperatures of all reservoirs are the same.\\

\proof of Proposition \ref{prop:affine}: 
We first note that $\tau_j>0$ implies $\rho^+_j>0$, thanks to Lemma \ref{lem:posrho}. Next we turn to the computation of $\sigma_j(\rho^+)$, $j\in\mathcal{J}$, according to (\ref{entsteady}).

For case i), the choice $\lambda_j=\gamma_j/\Gamma$ yields for (\ref{adjcur}) and (\ref{steadyj})
\begin{align}\label{expconv}
&\cL_j(\rho^+)=\gamma_j(\tau_j -\rho^+)\\
    &\rho_j^+=\Gamma \big(i[H,\cdot]+\Gamma\big)^{-1}(\tau_j),
\end{align}
so that (recall (\ref{recomb}))
\begin{align}\label{comrho}
\rho^+&=\sum_{j\in\mathcal{J}}\frac{\gamma_j}{\Gamma}\rho^+_j.
\end{align}

Also, (\ref{entprodcj}) yields
\begin{align}
\sigma_j(\rho^+)
    &=\gamma_j\tr \Big((\tau_j-T)(\ln(\rho^+_j)-\ln(\rho^+))\Big)\geq 0.
\end{align}
Consequently, by (\ref{entsteady}), or using the definition of $T$,
\begin{align}\label{formEP}
\sigma(\rho^+)=\sum_{j\in\mathcal{J}}\sigma_j(\rho^+)=\sum_{j\in\mathcal{J}}\gamma_j\tr \big((\tau_j-T)\ln(\rho^+_j)\big)\geq 0.
\end{align}

Now, the logarithm is operator concave, which means that, as operators,
\begin{align}
\sum_{j\in\mathcal{J}}\frac{\gamma_j}{\Gamma}\ln(\rho^+_j)\leq \ln \Big(\sum_{j\in\mathcal{J}}\frac{\gamma_j}{\Gamma}\rho^+_j \Big)\equiv \ln (\rho^+).
\end{align}
Since $T>0$, we have for any $A\geq 0$, $\tr (TA)=\tr (T^{1/2} A T^{1/2})\geq 0$, so that
\begin{align}
\sigma(\rho^+)\geq \sum_{j\in\mathcal{J}}\gamma_j(\tr \big(\tau_j\ln(\rho^+_j)\big)-\tr\big(T\ln(\rho^+)\big)).
\end{align}
By the second point of Lemma \ref{lem:posrho}, we get
\begin{align}
\sigma(\rho^+)&\geq 
\Gamma \Big(S(\rho^+)-\sum_{j\in\mathcal{J}} \frac{\gamma_j}{\Gamma} S(\rho^+_j)\Big),
\end{align}
where $S(\rho)$ is the entropy  of the state $\rho$. The entropy being strictly concave on $\mathcal{DM(H)}$, we deduce that
\be
\sigma(\rho^+)>0 \ \Leftarrow \ \exists \, j\neq k \ \mbox{s.t.}\ \rho_j^+\neq \rho^+_k 
\ \Leftrightarrow \ \exists \, j\neq k \ \mbox{s.t.}\ \tau_j\neq \tau_k.
\ee
Conversely, when all reset matrices are equal, we have for all $j\in \mathcal{J}$
\begin{align}
    \rho^+_j&=\gamma_j(\lambda_ji[H,\cdot]+\gamma_j)^{-1}(\tau), \ \ \rho^+=\Gamma(i[H,\cdot]+\Gamma)^{-1}(\tau),
\end{align}
and
\begin{align}
    \cL_j(\rho^+)=(\gamma_j-\Gamma \lambda_j)(\tau-\rho^+).
\end{align}
Since the prefactor vanishes for $\lambda_j=\gamma_j/\Gamma$,  $\sigma_j(\rho^+)=0$, for all $j\in \mathcal{J}$, hence the conclusion.

For case ii), we have on top of (\ref{sjlj})
\begin{align}
\rho_1^+=\gamma_1 \big(i[H,\cdot]+\gamma_1\big)^{-1}(\tau_1), \ \
    \cL_1(\rho^+)= \Gamma(\rho^+-T)+\gamma_1(\tau_1-\rho^+)
\end{align}
so that the non-negative individual entropy productions read 
\begin{align}\label{hamone}
    \sigma_1(\rho^+)&= \tr \Big((\Gamma(\rho^+-T)+\gamma_1(\tau_1-\rho^+))(\ln(\rho^+_1)-\ln(\rho^+))\Big)\nonumber\\
    \sigma_j(\rho^+)
    &=\gamma_j(S(\tau_j|\rho^+)+S(\rho^+|\tau_j)).
\end{align}
Thus
$\sigma_j(\rho^+)=0$ if and only if $\tau_j=\rho^+$, which proves the statement.

Now in case iii), $[H,\tau_j]=0$ for all $j\in\mathcal{J}$, then $\rho^+_j=\tau_j$, so that $[H,T]=0$ and $\rho^+=T$. Plugging this into (\ref{entprodcj}) yields
\begin{align}
    \sigma_j(\rho^+)=\tr \big(\gamma_j(\tau_j-\rho^+)(\ln(\tau_j)-\ln(\rho^+))\big),
\end{align}
from which the result follows.

\ep

\medskip

In the special case  where all reset matrices are identical, but {\bf DB} does not hold, we have a complete characterization of the strict positivity of the EP for convex (not merely affine) combinations of $\{\lambda_j\}_{\mathcal{J}}$. This corresponds to the situation where all dissipators are identical, {\it i.e.} all reservoirs have identical characteristics, however without the symmetry provided by {\bf DB}.  
\begin{prop}\label{prop:convex}  Assume {\bf Diss} and suppose $\tau_j=\tau>0$ for all $j\in\mathcal{J}$, with $[H,\tau]\neq 0$. Then, for any convex combination of $\{\lambda_j\}_{\mathcal{J}}$, $\sigma(\rho^+)\geq 0$ with equality if and only if $\lambda_j=\gamma_j/\Gamma$ for all $j\in\mathcal{J}$.
\end{prop}
\proof:
Our main assumption implies $\tau=T$ so that for all values of $\lambda_j$,
\begin{align}\label{fanciv}
    0\leq \sigma_j(\rho^+)=(\gamma_j-\lambda_j\Gamma)\big(S(\rho^+)-S(\rho_j^+)+S(\rho^+|\rho_j^+)\big),
\end{align}
as a consequence of (\ref{entprodcj}) and (\ref{useful}) applied twice.
This leads us to the following  generalization of (\ref{tbg}): 
\begin{lem}\label{defrll}
For $T>0$ consider
\be\label{defrl}
{\mathbb R}\ni \lambda\mapsto \rho(\lambda):=(\lambda i[H,\cdot]+1)^{-1}(T).
\ee 
If $[H,T]\neq 0$, the map ${\mathbb R}\ni \lambda\mapsto \rho(\lambda)$ is injective, and $\forall (\lambda, \mu)\in \mathbb R\times \mathbb R_* $,
    \begin{align}\label{ineqre}
        (1-{\lambda}/{\mu})(S(\rho(\mu))+ S(\rho(\mu)|\rho(\lambda))-S(\rho(\lambda)))\geq 0.
    \end{align}
    If, moreover, $0<\lambda/\mu < 1$, one has the strict inequality
    \begin{align}\label{strictineqre}
        S(\rho(\mu))+ S(\rho(\mu)|\rho(\lambda))-S(\rho(\lambda))> 0.
    \end{align}
\end{lem}
\begin{rem}
    In case $\mu=0$, $\rho(0)=T$ and the LHS of (\ref{strictineqre}) equal zero for all $\lambda$, by (\ref{tbg}), and for $\lambda=\mu \neq 0$, the same is true. However, for $\lambda=0$, thanks to (\ref{useful}), the LHS equals $S(T|\rho(\mu))+S(\rho(\mu)|T)>0$ for all $\mu\neq 0$, since $[H,T]\neq 0$. \\
    Inequality (\ref{fanciv}) corresponds to (\ref{ineqre}) with $\mu=1/\Gamma\in \R^+_*$, $\lambda=\lambda_j/\gamma_j\in \mathbb R$.\\
    Also, $\lim_{|\lambda|\rightarrow \infty}\rho(\lambda)=\sum_{j}P_jTP_j>0$, where $\{P_j\}$ are the spectral projectors of $H$, see (\ref{rpspecpro}).
    We already noted that $\lambda\mapsto \rho(\lambda)$ is meromorphic with simple poles at $\big\{\frac{i}{e_j-e_k}\big\}_{1\leq j, k \leq d}$. 
\end{rem}
\proof of Proposition \ref{prop:convex}: An immediate consequence of (\ref{fanciv}) and (\ref{strictineqre}), observing that $\lambda_j<\gamma_j/\Gamma$ must hold for one index $j$ at least if both convex combinations $\{\lambda_j\}_{\mathcal{J}}$, $\{\gamma_j/\Gamma\}_{\mathcal{J}}$ do not agree \ep\\

\proof of Lemma \ref{defrll}:
Injectivity of (\ref{defrl}) for $[H,T]\neq 0$ follows from the fact that $\rho(\lambda)=\rho(\mu)$ with $\lambda\neq \mu$ and $\mu\neq 0$,  implies
\be
T=(\lambda i[H,\cdot](\mu i[H,\cdot]+1)^{-1}(T)+\rho(\mu)\Rightarrow (\lambda/\mu -1)(T-\rho(\mu))=0.
\ee
But $\rho(\mu)=T$, with $\mu\neq 0$ is true only for $[H,T]=0$, which is excluded by assumption.

Now, for $\lambda\in \mathbb R$ and $\mu\neq 0$, set $\mathcal{L}_\lambda(\cdot)=-\lambda i[H,\cdot]+(T\tr(\cdot)-\cdot ) $ and consider
\be
-\frac{d}{dt}S(\e^{t\mathcal{L_\lambda}}(\rho(\mu))|\rho(\lambda))|_{t=0}=\tr \big(\mathcal{L}_\lambda(\rho(\mu))(\ln(\rho(\lambda))-\ln(\rho(\mu)))\big)\geq 0.
\ee
With
\be
\mathcal{L}_\lambda(\rho(\mu))=(1-{\lambda}/{\mu})(\rho(\mu)-T),
\ee
we get, using (\ref{useful}) twice,
\begin{align}
    &(1-{\lambda}/{\mu})\tr \big(T((\ln(\rho(\lambda))-\ln(\rho(\mu))))+\rho(\mu)(\ln(\rho(\mu))-\ln(\rho(\lambda)))\big)\nonumber\\
    &=(1-{\lambda}/{\mu})( S(\rho(\mu))+ S(\rho(\mu)|\rho(\lambda))-S(\rho(\lambda)))\geq 0,
\end{align}
which proves the first statement.
It thus remains to show (\ref{strictineqre}) for $0< \lambda /\mu <1$. Suppose
\be\label{absurd}
S(\rho(\mu))+ S(\rho(\mu)|\rho(\lambda))=S(\rho(\lambda))
\ee
for $0 < \lambda<\mu$. Then the prefactor of (\ref{ineqre}) with $\mu$ and $\lambda$ exchanged is negative since $\mu/\lambda >1$, so that we get, using (\ref{absurd}) 
\begin{align}
    0\geq S(\rho(\lambda))+ S(\rho(\lambda)|\rho(\mu))-S(\rho(\mu))=S(\rho(\mu)|\rho(\lambda))+ S(\rho(\lambda)|\rho(\mu)).
\end{align}
In turn, this imposes the non-negative relative entropies to be equal to zero, {\it i.e.} $\rho(\mu)=\rho(\lambda)$. By injectivity of the map $\rho$, this implies $\mu=\lambda$, a contradiction.
The same argument applies for $\lambda$ and $\mu$ both strictly negative and such that $0<\lambda/\mu<1$. \ep

Before treating the case of a tripartite structure, we illustrate and provide additional understanding of the behaviours of the entropy production for the case of a simple Hilbert space, considering an explicit model of a single qubit coupled to one reservoir and then to multiple reservoirs.

\section{EP of a \qrm for a single qubit: example}\label{exandnum}

In this section, we consider the case of a single qubit described by its Hamiltonian $H = \epsilon (\one + \sigma_z)/2 + \delta/2 \sigma_x$, with bare energy $\epsilon$ and tunneling  energy $\delta/2$. In the computational (canonical) basis $\{ \vert 0 \ket, \vert 1 \ket\}$, the Hamiltonian $H$ reads:
\begin{eqnarray}
    H = \left( \begin{array}{cc}
    0 & \delta/2 \\
   \delta/2 & \epsilon \end{array} \right)\,.
\end{eqnarray}
In contrast to the previous sections where results were written in the eigenbasis of the system, all results here will be provided in the computational basis. This choice allows us to provide additional insights into exact expressions illustrating the results derived in the previous sections.\\
As reset states, we consider diagonal states in the computational basis, of the form:
\begin{eqnarray}
\label{eq:reset}
    \tau_j =  \left( \begin{array}{cc}
     t_j & 0 \\
    0 & 1-t_j\end{array} \right)\,, \quad t_j \in ]0,1[\setminus \{1/2\}\,.
\end{eqnarray}
The condition on $t_j$ ensures $\tau_j>0$ and $\tau_j \neq \one$ to avoid trivialities.\\

With this class of reset states and given the form of $H$, we have the following equivalence:
\begin{eqnarray}
    [H, \tau_j] = 0 \Leftrightarrow \delta=0\,.
\end{eqnarray}

We also highlight that this choice of reset states implies that $[\tau_j, \tau_k]=0, \forall j,k$. These reset states correspond to a subset of reset states considered in the previous sections, and we note that this assumption could not be taken into account to derive additional results. Reset states of the form \eqref{eq:reset} can correspond to Gibbs states of thermal environments characterized by their inverse temperature $\beta_j$ if $\delta=0$ with
\begin{equation}
    \tau_j = \frac{e^{-\beta_j H}}{\tr(e^{-\beta_j H})} = \frac{e^{-\beta_j H}}{\mathcal{Z}}\,.
\end{equation}

In the following, we make explicit some of the previous results considering a single qubit coupled to one, two or three reservoirs. Its dynamics under a \qrm with reservoir $j$ is described by the Lindbladian that acts onto the density operator $\rho$ the system
\begin{equation}
    \cL_j \, \rho = -i [H, \rho] +  \gamma_j (\tau_j - \rho )\, \quad j=1, \ldots, N \,,
\end{equation}
with $\cL = \sum_j \cL_j$. For clarity, we express the density operator in the computational (canonical) basis $\{ \vert 0\ket, \vert 1 \ket \}$ and we will use indices $A,B, C$ instead of $j=1,\ldots, N$ for labelling up to 3 reset processes.

\subsection{Single qubit coupled to a single reservoir}

Here, we consider only a single \qrm with reset state $\tau_A$ and coupling constant $\gamma_A$. In the general case $[H, \tau_A] \neq 0$, the steady state is given by
\begin{eqnarray}
    \rho_+ =  \left( \begin{array}{cc}
\frac{\delta^2/2 +  (\epsilon^2+\gamma^2) t_A}{ \epsilon^2+\delta^2 + \gamma_A^2} & \frac{ \delta/2 \, (\epsilon + i \gamma_A) (1 - 2 t_A)}{\epsilon^2+\delta^2 + \gamma_A^2} \\
 \frac{ \delta/2 \, (\epsilon - i \gamma_A) (1 - 2 t_A)}{\epsilon^2+\delta^2 + \gamma_A^2} & 1 - \frac{\delta^2/2 +  (\epsilon^2+\gamma^2) t_A}{ \epsilon^2+\delta^2 + \gamma_A^2}
    \end{array} \right)\,.
\end{eqnarray}
It is straightforward to verify that $\rho_+ = \tau_A$ when $[H, \tau_A]=0$.\\

The exact solution at all times is given by (\ref{basisindep}), and when $[H, \tau_A]=0$, it reads
\begin{equation}
    \rho(t) = \left( \begin{array}{cc}
    e^{-\gamma_A t} p_{00}(0) + t_A (1-e^{-\gamma_A t})   & p_{01}(0) e^{(i \epsilon - \gamma_A) t}  \\
   p_{10}(0) e^{(-i \epsilon - \gamma_A) t} & e^{-\gamma_A t} p_{11}(0) + (1- t_A) (1-e^{-\gamma_A t}) \end{array} \right)\,.
\end{equation}
Here, $\{p_{00}(0), p_{01}(0), p_{10}(0), p_{11}(0)\}$ correspond to the matrix elements of the density operator at time $t=0$.\\

The entropy production at all times is given by \eqref{EPL}, where
\be
\sigma_\cL (\rho(t))=-\frac{d}{dt}S(\rho(t)|\rho^+), \ \ \mbox{with } \ \ \rho(t)=e^{t\cL}(\rho_0).
\ee
In case $[H, \tau_A]=0$ with the condition $p_{11}(0) = 1- p_{00}(0)$, we obtain:
\begin{eqnarray}
   \sigma_\cL (\rho(t)) &=& \gamma_A e^{-\gamma_A t} (p_{00}(0)- t_A)  \Bigg( \log \left( \frac{ p_{00}(0) e^{-\gamma_A t} -t_A (1-e^{-\gamma_A t})}{1- p_{00}(0) e^{- \gamma_A t} - t_A (1-e^{-\gamma_A t})} \right) \nonumber \\
   &&- \log\left( \frac{t_A}{1-t_A} \right)   \Bigg)\,. 
   \end{eqnarray}
In the long-time limit, $t \rightarrow \infty$, the EP vanishes as expected, $\rho_+ = \tau_A$ when $\delta =0$.

\subsection{Single qubit coupled to multiple reservoirs}

We now treat the case where the qubit is coupled to multiple reservoirs, allowing for non-equilibrium physics when $\tau_i \neq \tau_j$. We consider the case $N=3$ reservoirs, as the first non-trivial situation.\\

The \qrm for the dynamics of a single qubit coupled to three reservoirs labeled $A, B, C$ takes the form (considering $\tr(\rho)\equiv 1$):
\begin{eqnarray}
    \dot{\rho} = - i [H, \rho] + \gamma_A (\tau_A - \rho) + \gamma_B (\tau_B - \rho) + \gamma_C (\tau_C - \rho)\,.
\end{eqnarray}

\noindent In the computational basis of the qubit, the steady-state solution takes the simple form,
\begin{eqnarray}
    \rho_+ = \left( \begin{array}{cc}
     \frac{(\epsilon^2 + \Gamma^2) \bar{t} +  \delta^2/2 }{ (\epsilon^2 + \delta^2 + \Gamma^2)} & \frac{\delta/2 \, (\epsilon - i \Gamma) (1 - 2 \bar{t})}{\epsilon^2 + \delta^2 + \Gamma^2} \\
      & \\
     \frac{\delta/2 \, (\epsilon + i \Gamma) (1 - 2 \bar{t})}{\epsilon^2 + \delta^2 + \Gamma^2} & 1 - \frac{(\epsilon^2 + \Gamma^2) \bar{t} +  \delta^2/2 }{ (\epsilon^2 + \delta^2 + \Gamma^2)} \end{array} \right).
\end{eqnarray}
with 
\be \Gamma = \gamma_A + \gamma_B + \gamma_C \ \ \mbox{and}  \ \ \bar{t} := \sum_{j\in\{A,B,C\}}\ \gamma_j t_j / \Gamma,
\ee 
the average ground state population of the density operator $T$ (\ref{recomb}). Let us note that the case of a single qubit coupled to two reservoirs is simply obtained by imposing $\gamma_C=0$  in the above expression and $\gamma_C = \lambda_C=0$ when introducing the affine coefficients below. When $[H, \tau_j] =0 \, \forall j\in\{A,B,C\}$, we recover the steady-state solution $\rho^+ = T$.\\

The propositions and lemmas in the previous sections were demonstrated using specific affine combinations considering the Lindbladian:
\begin{eqnarray}
   \cL(\rho) = \sum_j \cL_j(\rho) = \sum_j \, - i [\lambda_j H, \rho] + \gamma_j (\tau_j - \rho )\,,
\end{eqnarray}
with real parameters $\lambda_j$ satisfying $\sum \lambda_j =1$. For the case of three reservoirs, the individual steady states $\rho_j^+$ for a given $\lambda_j$ are solutions of $\cL_j(\rho_j^+)=0$, $j\in\{A,B,C\}$,
\begin{eqnarray}
    \rho_j^+ = \left( \begin{array}{cc}
    \frac{\gamma_j^2 t_j + \lambda_j^2 (\epsilon^2 t_j + \delta^2/2)}{\gamma_j^2 + \lambda_j^2 (\epsilon^2+ \delta^2)} & \frac{\lambda_j \, \delta/2 \, (\epsilon \lambda_j - i \gamma_j)(1 - 2 t_j)}{\Gamma_j^2 + \lambda_j^2(\epsilon^2 + \delta^2)} \\
     & \\
    \frac{\lambda_j \, \delta/2 \, (\epsilon \lambda_j + i \gamma_j)(1 - 2 t_j)}{\Gamma_j^2 + \lambda_j^2(\epsilon^2 + \delta^2)} & 1-  \frac{\gamma_j^2 t_j + \lambda_j^2 (\epsilon^2 t_j + \delta^2/2)}{\gamma_j^2 + \lambda_j^2 (\epsilon^2+ \delta^2)} \end{array} \right) \,. 
\end{eqnarray}

\noindent The steady-state EP $\sigma (\rho^+)$ is the sum of the individual EP $\sigma_j(\rho^+)$
\begin{eqnarray}
  \sigma(\rho^+) = \sum_j \sigma_{j\in\{A,B,C\}}(\rho^+)\,,
  \end{eqnarray}
with the individual EP $\sigma_j$ defined as:
\begin{eqnarray}
    \sigma_j (\rho^+) = \tr \left( \cL_j(\rho^+) \Big( \log (\rho_j^+) - \log(\rho^+) \Big) \right)\,.
\end{eqnarray}
In case $\tau_j=\tau$ for all $j\in\{A,B,C\}$, the expression reduces to:
\begin{eqnarray}
    \sigma_j(\rho^+) = \frac{ \gamma_j (-1 + \lambda_j \Gamma)^2 \, \kappa_j(\lambda_j) \, \delta^2/2  \,\left( \log\left( 1 + \kappa_j(\lambda_j) \right) - \log\left( 1- \kappa_j(\lambda_j) \right) \right) }{(\epsilon^2 \lambda_j^2 + \gamma_j^2) (\epsilon^2 + \Gamma^2 + \delta^2)}\,,
\end{eqnarray}
with, for clarity, the definition
\begin{eqnarray}
    \kappa_j(\lambda_j) = \sqrt{\frac{(\epsilon^2 \lambda_j^2 + \gamma_j^2) (1- 2t_j)^2}{\epsilon^2 \lambda_j^2 + \gamma_j^2 + \lambda_j^2 \delta^2}}\,.
\end{eqnarray}
When $\delta=0$, it is straightforward to see that the numerator of the individual EPs vanishes, implying $\sigma_j(\rho^+) =0$ when $[H, \tau_j]=0$. \\

\noindent We now illustrate {\bf Propositions 4.3} and {\bf 4.5.}, first investigating the total EP $\sigma(\rho^+)$ as a function of the ground-state populations $t_A$ and $t_B$ characterizing the reset states (panels a) and b)), and then, as a function of the affine parameters $\lambda_A, \lambda_B, \lambda_C$ (panels c) and d)). In panel a), we show that the total EP, for fixed values of $\lambda_j= \gamma_j/\Gamma$, is always positive, and vanishes only when $\tau_A = \tau_B$. We also note that the total EP does not exhibit specific symmetries, as expected for non specific values for coupling parameters. In panel b), we provide a zoom on the region in the vicinity of $t_A = t_B$: the quadratic behaviour of the EP close to its zero value reflects a generic behaviour of the EP close to zero. In panel c), we show the total EP as a function of $\lambda_A$ and $\lambda_B$, fixing $\lambda_C = 1-(\lambda_A + \lambda_B)$. For the three reset states being the same, $\tau_A= \tau_B = \tau_C$, the total EP is again always positive, and only vanishes when $\lambda_A= \gamma_A/\Gamma$ and $\lambda_B = \gamma_B / \Gamma$, which illustrates {\bf Proposition 4.5}. The same quadratic behavior of the total EP close to its zero values is observed, see zoom in panel d). \\

  \begin{figure*}
        \centering
        \includegraphics[width=1.0\textwidth]{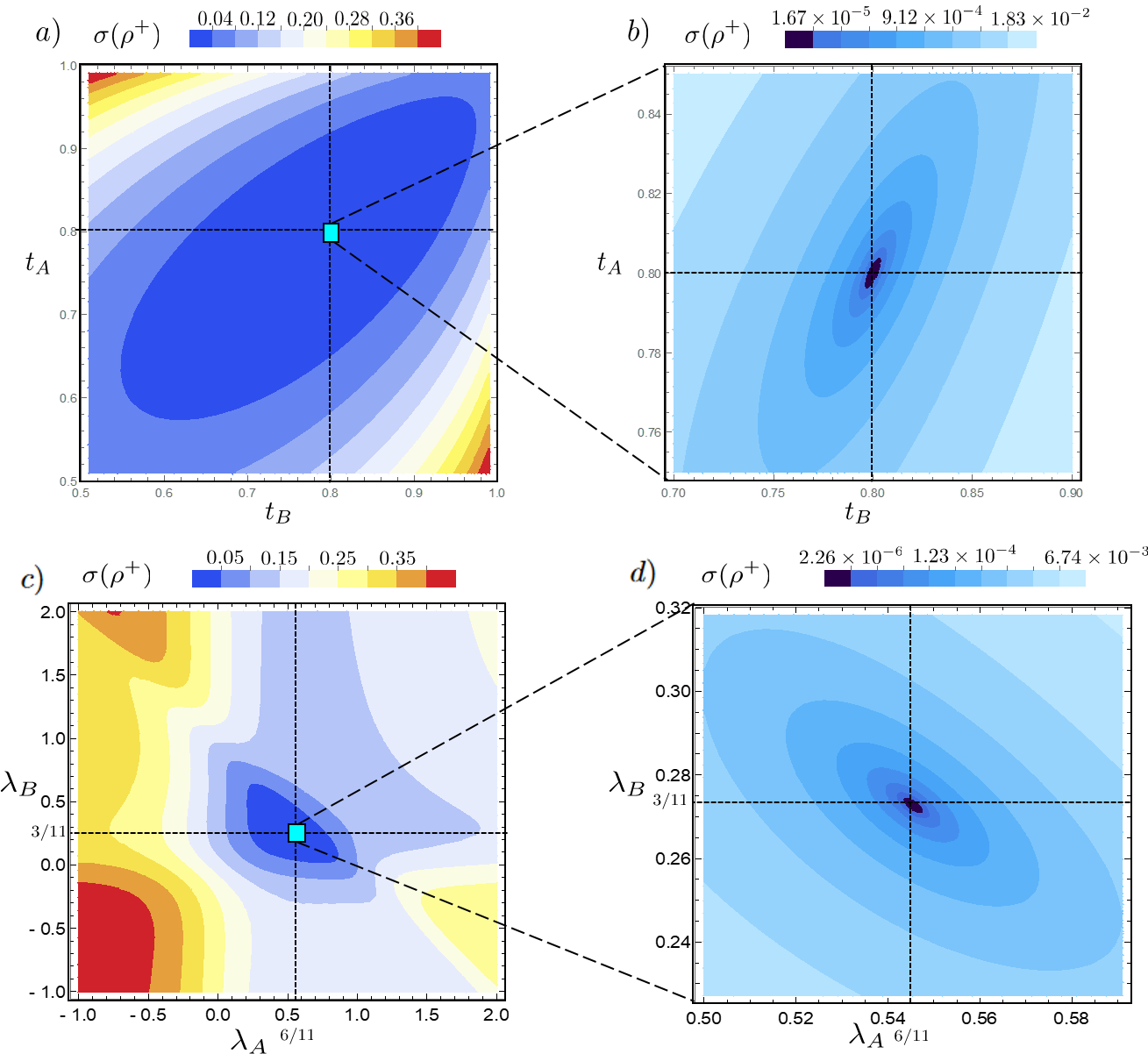}
        \caption{Illustration of {\bf Propositions 4.3} and {\bf4.5}. Panels a) and b): Contour plots of the total EP $\sigma (\rho^+)$ as a function of $t_A, t_B$, with $\lambda_j = \gamma_j/\Gamma$. Numerical values for the different parameters in units of $\epsilon$ are fixed: $\delta=0.7, \gamma_A=1, \gamma_B=1/2, \gamma_C = 1/3, t_C =0.8$.  Panels c) and d): Contour plots of the total EP $\sigma (\rho^+)$ as a function of $\lambda_A, \lambda_B$, with $\tau_A = \tau_B = \tau_C$. The total EP vanishes only when $\lambda_j = \gamma_j/\Gamma$, explicitly when $\lambda_A = 6/11, \lambda_B = 3/11$ (marked by the two intersecting dashed lines). $\sigma(\rho_+) \geq 0$ in the whole range for affine parameters, including $\lambda_A, \lambda_B <0$, a case which could not be addressed analytically.  Numerical values for the different parameters in units of $\epsilon$ are fixed: $\delta=0.7, \gamma_A=1, \gamma_B=1/2, \gamma_C = 1/3, t_A = t_B=t_C =0.9$.}
        \label{fig:43_v2}
    \end{figure*}

     \begin{figure*}
        \centering
        \includegraphics[width=1.0\textwidth]{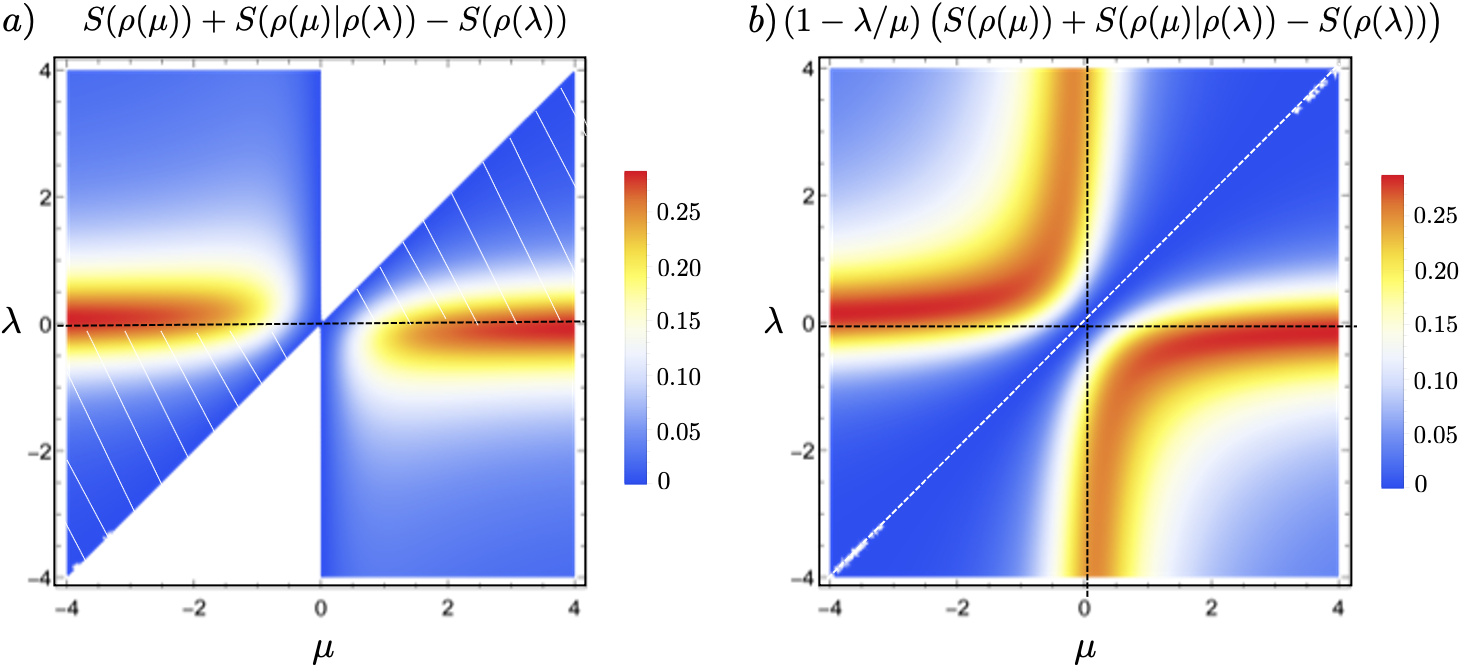}
        \caption{Role of affine parameters for Lemma 4.6. Numerical values of the parameters in units of $\epsilon$: $\delta =0.7, \gamma_A=1, t_A =0.9$. Panel a): Density plot of the function $S(\rho(\mu)) + S(\rho(\mu)\vert\rho(\lambda))- S(\rho(\lambda))$ as a function of $\lambda, \mu$ in the range $[-4,4]$.  Panel b): Illustration of Eq.~\eqref{ineqre}, showing positivity of $(1- \lambda/\mu)S(\rho(\mu)) + S(\rho(\mu)\vert\rho(\lambda))- S(\rho(\lambda))$ for all $\lambda, \mu$. Dashed white line corresponds to $\mu = \lambda$, values for which this quantity vanishes.
        }
        \label{fig:46}
    \end{figure*}

\noindent Point ii) of {\bf Proposition 4.3.} can be made explicit through the following example. Let us consider $\lambda_A =1, \lambda_B = \lambda_C =0$. When the three reset states are the same $\tau_A = \tau_B = \tau_C$, the total EP reduces to
    \begin{eqnarray}
    &&  \sigma(\rho_+) = \frac{(\gamma_B + \gamma_C) \, \delta^2/2 \, (1-2 t_A) }{(\epsilon^2 + \gamma_A^2 + \delta^2 ) (\epsilon^2 + \Gamma^2 + \delta^2) \kappa_A(1)} \cdot \nonumber \\
      && \Bigg[ \gamma_A (\gamma_B + \gamma_C) (1- 2 t_A) \Big( \log\left(  \frac{1}{2}+ \kappa_A (1)/2 \right) - \log\left(  \frac{1}{2} - \kappa_A (1)/2 \right) \Big) \nonumber \\
      && + (\epsilon^2 + \gamma_A^2 + \delta^2) \kappa_A(1) \log \left( \frac{1-t_A}{t_A} \right)\Bigg]\,.
    \end{eqnarray}
From this expression, it is again straightforward to see that the total EP vanishes when $[H, \tau_A]=0$, corresponding to $\delta=0$.
    
Similarly, if we now assume $[H, \tau_j]=0 \, \forall j\in\{A,B,C\}$, i.e. $\delta=0$, the individual EP takes the simple form:
    \begin{eqnarray}
        \sigma_j (\rho_+) &=& \gamma_j (-t_j + \bar{t}) \left[ \log (1-t_j) - \log(1- \bar{t})\right] \nonumber 
        + \gamma_j (t_j - \bar{t}) \left[ \log (t_j) - \log( \bar{t})\right]\,,
    \end{eqnarray}
    which corresponds to the sum of the two relative entropies:
    \begin{eqnarray}
        \sigma_j (\rho_+) = \gamma_j \big( S(\tau_j \vert T) + S(T \vert \tau_j) \big)\,.
    \end{eqnarray}
It becomes straightforward to see that the individual EP vanishes only when $\tau_j = T$ $\forall \, j\in\{A,B,C\}$. We can also easily verify that the total EP vanishes if $\lambda_j = \bar{t} = \sum_j \gamma_j t_j/\Gamma$ $\forall j\in\{A,B,C\}$. This provides explicit expressions for Point iii) of {\bf Proposition 4.3}. \\

Figure~\ref{fig:46} numerically investigates {\bf Lemma 4.6.}. Equation~\eqref{strictineqre} corresponds to the two regions dashed with white lines and we confirm the positivity with the density plot in panel a). White regions correspond to negative values of $S(\rho(\mu)) + S(\rho(\mu)\vert\rho(\lambda))- S(\rho(\lambda))$, when $\lambda/\mu>1$. In addition to the analytical results, numerical calculations show that this quantity is also positive for all $\lambda, \mu$ when $\lambda/\mu <0$.

\section{\qrm  on Tri-partite Hilbert spaces}\label{sec:struct}

The setup here is the one we developed in \cite{HJ} to analyse the \qrm dynamics of tri-partite systems. This structure has the merit of making distinct the description of the reservoirs and that of the quantum system interacting with the environment they form. Our goal is to estimate the EP within this structured framework that we recall, mainly in order to set the notation.

We consider a structured Hilbert space $\cH=\cH_A\otimes\cH_C\otimes\cH_B$, where $\cH_\#$ are Hilbert spaces, of dimensions noted $n_\#<\infty$, where $\#\in\{A,B,C\}$. Two reset matrices $\tau_A\in {\cal DM(H}_A)$, $\tau_B\in {\cal DM(H}_B)$ are defined on their respective Hilbert space and $\gamma_A, \gamma_B>0$ are the corresponding positive rates. Let $H_A, H_B, H_C$ be three Hamiltonians acting on their respective Hilbert space.

\medskip

Let us first define the Lindblad operator in the uncoupled case ({\it i.e. } when the system $A-C-B$ is non-interacting), and we introduce and analyse below the effect of adding a small Hamiltonian coupling between the parts of the system  $A-C-B$.

\medskip

Let the {\it uncoupled} generator of the \qrm be 
\begin{align}\label{triqrm}
\cL(\rho)=&-i[H_A\otimes \un_{C}\otimes \un_{B}+   \un_{A}\otimes H_C\otimes \un_{B}  + \un_{A}\otimes \un_{C}\otimes H_B ,\rho]\\ \nonumber
&+\gamma_A(\tau_A\otimes \tr_A(\rho)-\rho)+\gamma_B(\tr_B(\rho)\otimes \tau_B-\rho),
\end{align}
where $\un_{\#}$ denotes the identity operator on $\cH_\#$
and $\tr_\#$ denotes the operator on the tensor product of Hilbert spaces with indices different from $\#$, obtained by taking the partial trace over $\cH_\#$. Similarly, we denote by 
$\tr_{\#\#'}$  the operator on the Hilbert space with index different from $\#$ and $\#'$ obtained by taking the partial trace over $\cH_\#\otimes\cH_{\#'}$. For example,
\be
\tr_A : \cB(\cH_A\otimes\cH_C\otimes\cH_B)\ra \cB(\cH_C\otimes\cH_B),\ \  \tr_{AB} : \cB(\cH_A\otimes\cH_C\otimes\cH_B)\ra \cB(\cH_C)
\ee
will be viewed as linear maps.
We write $H_\#$ for the Hamiltonian both on $\cH_\#$ and $\cH$, the context making it clear what is meant. Finally, we denote the dissipator part of the generator by 
\begin{align}\label{dissip}
\cD(\rho)&=\cD_A(\rho)+\cD_B(\rho)\nonumber\\
&=\gamma_A(\tau_A\otimes \tr_A(\rho)-\rho)+\gamma_B(\tr_B(\rho)\otimes \tau_B-\rho),
\end{align}
so that 
$
\cL(\rho)=-i\big[H_A+H_C  +H_B , \rho\big]+\cD_A(\rho)+\cD_B(\rho).
$
\medskip

For $g\in \R^*$ a coupling constant and $H=H^*\in \mathcal{B(H)}$, we define the Lindblad generator of the coupled \qrm acting on $\mathcal{B(H)}$ by
\be\label{couqrmo}
\cL_g(\rho)=\cL(\rho)-ig[H,\rho]
\ee
with $\cH=\cH_A\otimes\cH_C\otimes\cH_B$. Here, $H$
is a Hamiltonian that effectively couples the different Hilbert spaces $\cH_\#$, while $g\in \R$ measures the strength of the coupling.

\section{Small Hamiltonian drive regime}\label{sec:smalldrive}

In the present paper, we will consider that the Hamiltonian part of the Lindblad generator which drives the system is small, {\it i.e.}, of order $g\ra 0$, as in Section 6 of \cite{HJ}. This means we consider 
\begin{align}\label{smalldrive}
    H_\#=0, \ \ \#\in \{A, B, C\},
\end{align}
so that 
\begin{align}\label{couqrm}
&\cL_g(\cdot)=-ig[H,\cdot]+\cD_A(\cdot)+\cD_B(\cdot)\equiv \cL_0(\cdot)+g\cL_1(\cdot), \ {\rm where}\nonumber\\
& \cL_0(\cdot)=\cD_A(\cdot)+\cD_B(\cdot), \ \ \cL_1(\cdot)=-i[H,\cdot].
\end{align}

As in the previous sections, we distribute the Hamiltonian parts of the generator among the two pieces of the environment that we will call the partial Lindbladians: for $\lambda\in \R$, we set 
\begin{align}\label{decomp3}
&\cL_g(\cdot)=\cL_{\lambda g}^{A}(\cdot)+\cL_{(1-\lambda) g}^{B}(\cdot), \ \ {\rm where}\\
 &   \cL_{\lambda g}^A(\cdot)=-ig\lambda[ H,\cdot]+\cD_A(\cdot), \ \ \cL_{(1-\lambda)g}^B(\cdot)=-ig(1-\lambda)[ H,\cdot]+\cD_B(\cdot).
\end{align}
The notation reflects the fact that doing so amounts to modifying the coupling constants.\\

We will make precise assumptions below that ensure $\cL_g(\cdot)$, $\cL_{g}^A(\cdot)$ and  $\cL_{g}^B(\cdot)$ all admit a unique faithful stationary state $\rho_g^+>0$, $\rho_g^A>0$, $\rho^B_g>0$, for $g$ small enough. This forbids us to consider $\lambda\in\{0,1\}$, since one of the partial Lindbladians $\cL_g^\#$ would act non trivially on $\cH_\#$ only which implies a large degeneracy of its kernel, see \cite{HJ}. Hence from now on, $\lambda\in\{0,1\}^{\rm C}$. 

\medskip

According to section \ref{genlindframe} and (\ref{decomp3}), the EP of the QDS $(\e^{t\cL_g})_{t\geq 0}$ in the steady state $\rho_g^+>0$ is given by
\begin{eqnarray}
 0\leq &\sigma(\rho_g^+)&=\sigma_A(\rho_g^+)+\sigma_B(\rho_g^+)\\ \nonumber
    &=&\tr\Big(\cL_{\lambda g}^A(\rho_g^+)\big(\ln(\rho_{\lambda g}^A)-\ln(\rho_g^+)\big)\Big)
    +\tr\Big(\cL_{(1-\lambda) g}^B(\rho_g^+)\big(\ln(\rho_{(1-\lambda) g}^B)-\ln(\rho_g^+)\big)\Big)\nonumber \\ 
    &=&\tr\big(\cL_{\lambda g}^A(\rho_g^+)\ln(\rho_{\lambda g}^A)\big)
    +\tr\big(\cL_{(1-\lambda) g}^B(\rho_g^+)\ln(\rho_{(1-\lambda) g}^B)\big), \label{entropy_flux}
\end{eqnarray}
where $\sigma_\#(\rho_g^+)$, $\#\in\{A,B\}$, defined by the middle line are non negative, while the entropy fluxes of the last line are not necessarily positive.\medskip

We shall conduct the analysis of $\sigma_g(\rho_g^+)$ to lowest order in the perturbative regime $g\ra 0$. 
Theorem 5.2 in \cite{HJ} states that  $\rho_g^+$ is analytic in $g$, for $|g|<g_0$, $0<g_0$ small, and provides expressions for the expansion of $\rho_g^+$  in powers of $g$ under some genericity hypotheses we now specify. This requires a few more definitions.

\medskip

Let the self-adjoint operator on $\cB(\cH_C)$
\begin{align}\label{hbar}
\overline{H}^{\, \tau} &=  \tr_{AB}(\tau_A^{1/2}\otimes \un_C \otimes\tau_B^{1/2}\, H\, \tau_A^{1/2}\otimes \un_C \otimes\tau_B^{1/2})\\ \nonumber
&=\tr_{AB}(H\, \tau_A\otimes \un_C \otimes\tau_B)=\tr_{AB}(\tau_A\otimes \un_C \otimes\tau_B \,H).
\end{align}
We shall assume

\medskip

\noindent
{\bf Spec($\overline{H}^{\, \tau}$)}:\\
\noindent 
The spectrum of $\overline{H}^{\, \tau}\in\cB(\cH_C)$ is simple, $\spec (\overline{H}^{\, \tau})=\{e_j^\tau\}_{1\leq j\leq n_C}$ and the corresponding Bohr frequencies 
$\{e_j^\tau-e_k^\tau\}_{1\leq j\neq k\leq n_C}$ are distinct. 

\medskip

The normalised eigenvectors of $\overline{H}^{\, \tau}$ are denoted by $\ffi_j^\tau$ and we let $\diag_\tau$ be the projector that extracts the diagonal part of matrices on $\cH_{\cC}$ in the orthonormal eigenbasis $\{\ffi_j^\tau\}$. \medskip

Define
$\Phi(\cdot):\cB(\cH_C)\ra \cB(\cH_C)\cap \{\rho_C\, | \, \tr \rho_C=~0\}$ by
\begin{align}\label{phi}
\Phi(\cdot)&= \tr_{AB}\big(\big[H, \cL_0^{-1}([H, \tau_A\otimes \diag_\tau(\, \cdot \,) \otimes \tau_B])\big]\big) \ \mbox{and}\\
\Phi_D(\cdot)&=\diag_\tau\,  \Phi(\cdot) \,  |_{\diag_\tau \, \cB(\cH_C)}. \label{deffid}
\end{align}
It is shown in \cite{HJ}, Section 6, that the dissipator $\cL_0$  (\ref{couqrm}) is indeed invertible on its argument in (\ref{phi}). More precisely:
for $\tilde \rho_0\in \cB(\cH)$ such that $\tr_{AB}(\tilde \rho_0)=0$, 
\begin{align}\label{dissinv}
\cL_0^{-1}(\tilde \rho_0)=\frac{-1}{\gamma_A+\gamma_B}\left\{  \tilde \rho_0+\frac{\gamma_A}{\gamma_B} \tau_A \otimes \tr_A(\tilde \rho_0)+\frac{\gamma_B}{\gamma_A}\tr_B(\tilde \rho_0)\otimes \tau_B\right\}.
\end{align}

The next assumption states that the coupling induced by $H$ is efficient. \medskip

\noindent
{\bf Coup}:\\
The linear map $\Phi_D (\cdot)$ on $\diag_\tau \, \cB(\cH_C)$ is s.t. $\dim \ker\, \Phi_D=1$.
\medskip

Actually, {\bf Coup} is equivalent to the statement
$\Phi_D^{-1}$ exists on the 
subspace $ \diag_\tau \, \cB(\cH_C)\cap \{\rho_C\, | \, \tr \rho_C=~0\}=~\ran \Phi_D$. Moreover, Section 6 of \cite{HJ} provides an explicit matrix representation of $\Phi_D$. 
\begin{rem}\label{remerr}
     We take the opportunity to correct a statement in Proposition 6.1 of \cite{HJ} regarding the validity of {\bf Coup}. The correct statement is  that the criterion in Proposition 6.1 is only sufficient for {\bf Coup} to hold (not necessary and sufficient). Details are provided in Appendix.
\end{rem}

Under {\bf Spec($\overline{H}^{\, \tau}$)} and {\bf Coup}, Theorem 5.2 of \cite{HJ} implies in particular that 
\begin{align}\label{exprho+}
    \rho_g^+&=\rho_0+g\rho_1+O(g^2)
\end{align}
with 
\begin{align}\label{rhozero}
\rho_0&=\tau_A\otimes \rho_0^C\otimes \tau_B\in \mathcal{DM(H)}, \ \mbox{where} \ \rho_0^C\in \ker \Phi_D\cap \mathcal{DM(H)}, \nonumber \\ 
\rho_1&=R_1+\tau_A\otimes r_C^{(1)}\otimes \tau_B,
\end{align}
where
\begin{align}\label{expfirstcorr}
&R_1=i\cL_0^{-1}([H, \rho_{0}]), 
\\ \nonumber
&\offdiag_\tau r_C^{(1)}=-i[\overline{H}^{\, \tau},  \cdot  ]^{-1}\Big( \offdiag_\tau \tr_{AB}\big(\big[H, \cL_0^{-1}([H, \rho_{0}])\big]\big)\Big),
\\ 
&\diag_\tau r_C^{(1)}=-\Phi_D^{-1}\big(\diag_\tau \tr_{AB}([H,\cL_0^{-1}(i[H, R_{1}+\tau_A\otimes\offdiag_\tau r_C^{(1)}\otimes\tau_B])])\big).\label{offdiagrj}
\end{align}

Similar definitions and criteria hold in order to ensure $\rho^\#_g$ admit expansions in powers of $g$, for $\#\in\{A,B\}$, see also Remark 3.1 in \cite{HJ}. \\

Let 
\begin{align} \label{Htausharp}
\overline{H}^{\, \tau_A}&
=\tr_{A}((\tau_A\otimes \un_C \otimes \un_B)H)\in \cB(\cH_C\otimes \cH_B),\\
\overline{H}^{\, \tau_B}&
=\tr_{B}((\un_A\otimes \un_C \otimes \tau_B)H)\in \cB(\cH_A\otimes \cH_C).
\end{align}
We assume 

\medskip 

{\bf Spec ($\overline{H}^{\, \tau_\#}$)}: The spectrum of $\overline{H}^{\, \tau_\#}$ is simple, with distinct Bohr frequencies. 

The normalised eigenvectors of $\overline{H}^{\, \tau^\#}$ are denoted by $\ffi_j^{\tau^\#}$ and we let $\diag_{\tau^\#}$ be the projector that extracts the diagonal part of matrices on $\cH_{C}\otimes \cH_{B}$, resp. $\cH_{A}\otimes \cH_{C}$, for $\#=A$, resp. $\#=B$,  in the orthonormal eigenbasis $\{\ffi_j^\tau\}$. \medskip

For $\#=A$, we set
$\Phi^A(\cdot):\cB(\cH_C\otimes \cH_B)\ra \cB(\cH_C\otimes \cH_B)\cap \{\rho_{CB}\, | \, \tr \rho_{CB}=~0\}$ by
\begin{align}\label{phiA}
\Phi^A(\cdot)&= \tr_{A}\big(\big[H, \cD_A^{-1}([H, \tau_A\otimes \diag_{\tau^A}(\, \cdot \,)])\big]\big) \ \mbox{and}\\
\Phi^A_D(\cdot)&=\diag_{\tau^A}\,  \Phi^A(\cdot) \,  |_{\diag_{\tau^A} \, \cB(\cH_C\otimes \cH_B)}, \label{deffidA}
\end{align}
and we define $\Phi^B(\cdot), \Phi^B_D(\cdot)$ analogously, {\it mutatis mutandis}.

Then we introduce the corresponding conditions of the efficiency of the coupling for $\#\in \{A,B\}$.\\

\noindent
{\bf Coup$^\#$}:\\
The linear map $\Phi_D^\# (\cdot)$ on its natural domain is s.t. $\dim \ker\, \Phi_D^\#=1$.
\medskip

Applying Theorem 5.2 of \cite{HJ} again, we get under {\bf Spec ($\overline{H}^{\, \tau_\#}$)} and {\bf Coup$^\#$}

\begin{align}\label{exprhosharp}
    \rho_g^\#&=\rho_0^\#+g\rho_1^\#+O(g^2),
\end{align}
for $\#\in\{A,B\}$, where
\begin{align}
    \rho_0^A=\tau_A\otimes \rho_0^{CB}, \ \ \rho_0^B=\rho_0^{AC}\otimes \tau_B,
\end{align}
and $\rho_0^{CB}$, $\rho_0^{AC}$ are states uniquely specified by $\rho_0^\#\in \ker \Phi_D^\# (\cdot)$, and first order corrections determined by the expressions (\ref{rhozero}), (\ref{expfirstcorr}) adapted to the context $\#\in\{A,B\}$.

We need to assume at this point that the leading orders of these expansions are positive definite.\\

\noindent
{\bf Pos}: \\
The reset matrices $\tau_A, \tau_B$ as well as the states $\rho_0^C, \rho_0^{BC}$ and $\rho_0^{AC}$ defined on their respective Hilbert spaces are all positive definite. \medskip

At this point, we can state the main Theorem of this Section. 
\begin{thm}\label{thmmain}
    Assume hypotheses 
    {\bf Spec ($\overline{H}$)}, {\bf Spec ($\overline{H}^{\, \tau_\#}$)}, {\bf Coup}, {\bf Coup$^\#$}, and {\bf Pos.}, and consider the tri-partite \qrm Lindbladian  $\cL_g=\cL_{\lambda g}^A+\cL_{(1-\lambda)g}^{B}$ for $\lambda\in \R\setminus \{0,1\}$. There exists $g_0(\lambda)>0$ small enough such that $\sigma(\rho_g^+)\geq 0,$ the total EP of the asymptotic state $\rho_g^+$ for QDS $(\e^{t\cL_{g} })_{t\geq 0}$, satisfies for all $0<|g|<g_0(\lambda)$ small enough $$\sigma(\rho_g^+)=g^2 \sigma^{(2)}(\lambda)+O_\lambda(g^3),$$ 
    where either $ \sigma^{(2)}(\lambda)>0$ for all $\lambda\in \R\setminus \{0,1\}$, with the possible exception $\sigma^{(2)}(\lambda_0)\geq 0$ for a unique $\lambda_0\in \R$, or  $\sigma^{(2)}\equiv 0$ and $\sigma(\rho_g^+)=O_\lambda(g^4)$.
\end{thm}
\begin{rem}
    The error term being $O(g^4)$ in case $\sigma^{(2)}\equiv 0$ stems from $\sigma(\rho_g^+)\geq 0$ for all $|g|$ small enough.
\end{rem}

We also provide a simple criterion ensuring $\sigma^{(2)}\not\equiv 0$.

\begin{prop}\label{prop:critpot}
    Under the hypotheses of Theorem \ref{thmmain}, we have
    \begin{align}\label{critpot}
        [H,\rho_0]\neq 0 \ \Leftrightarrow \  \sigma^{(2)}\not\equiv 0, \ \forall \lambda \in \R\setminus \{0,1\}.
    \end{align}
\end{prop}
\begin{rem}
    We emphasize here that we will treat explicitly a specific model of a tri-partite QRM in the next section, illustrating and testing the above theorems and propositions. Since some of the quantities of interest to do so are introduced in the proofs below, we proceed with the proofs, deferring the example to the next section.
\end{rem}

\proof of Theorem \ref{thmmain}:
 Assumption {\bf Pos.} allows us in turn to get expansions for the logarithms of the asymptotic states by functional calculus. We simply state the result here, referring the reader to Lemma B.1 of \cite{HJPR} for a proof.
\begin{lem} Let $\cH$ be a finite dimensional Hilbert space and 
$0<\rho\in \cB(\cH)$ with spectral decomposition 
$$\rho=\sum_{j\in \mathcal{I}} r_j P_j, \ \ r_j>0, \ P_j=P_j^2=P_j^*,$$
where $\mathcal{I}$ is a finite set of indices. 
Let $\Delta\in \cB(\cH)$ s.t. $\|\Delta\|<\min_{j\in \mathcal{I}} (r_j)$ and consider $\tilde \rho=\rho+\Delta$.
Then for the analytic function $\ln:\C\setminus \R^+$ it holds
\begin{align}\label{deflogR}
    \ln(\tilde \rho)=&\ln(\rho)+\frac{1}{2i\pi}\int_C(\rho-z)^{-1}\Delta (\rho-z)^{-1} \ln(z)dz+O(\|\Delta\|^2)\nonumber\\
    =&\ln(\rho)+\left\{\sum_j P_j\Delta P_j\frac{1}{r_j}+\sum_{i<j}(P_i\Delta P_j+P_j\Delta P_i)\frac{\ln(r_i)-\ln(r_j)}{r_i-r_j}\right\}\nonumber\\
    &+O(\|\Delta\|^2),
\end{align}
where $C$ is a positively oriented simple path in $\C\setminus \R_-$ encircling the spectrum of $\tilde \rho$.  
\end{lem}
Applying this lemma to the expansions (\ref{exprho+}) and (\ref{exprhosharp}), we get 
\begin{align}\label{explog+}
    \ln(\rho_g^+)&=\ln(\rho_0)+g Q_1+O(g^2), \\
\label{explogsharp}
    \ln(\rho_g^\#)&=\ln(\rho_0^\#)+g Q_1^\#+O(g^2),
\end{align}
where $Q_1$, resp. $Q_1^\#$ are obtained from (\ref{deflogR}) with $\rho_1$, resp. $\rho_1^\#$ in place of $\Delta$.

These expansions allow us to derive the leading order expressions for the individual EP:
\begin{prop}\label{prop:secordr}
    Under hypothese {\bf Spec ($\overline{H}$)}, {\bf Spec ($\overline{H}^{\, \tau_\#}$)}, {\bf Coup}, {\bf Coup$^\#$}, and {\bf Pos.}, for any fixed $\lambda\in \{0,1\}^{\rm C}$, and for $0<|g|\ra 0$,
    \begin{align}
        0\leq \sigma_A(\rho^+_g)=&g^2\, \tr \Big(-\lambda^2 i[H,\rho_0]Q_1^A\nonumber\\
        &+\lambda(\cD_A(\rho_1)Q_1^A+i[H,\rho_0]Q_1-i[H,\rho_1](\ln(\rho_0^A)-\ln(\rho_0)) \label{expsigmaA+} \nonumber\\
    &-\cD_A(\rho_1)Q_1\Big)+O_\lambda(g^3)\\
    0\leq \sigma_B(\rho^+_g)
    =&g^2\, \tr \Big(-(1-\lambda)^2 i[H,\rho_0]Q_1^B\nonumber\\
    &+(1-\lambda)(\cD_B(\rho_1)Q_1^B+i[H,\rho_0]Q_1-i[H,\rho_1](\ln(\rho_0^B)-\ln(\rho_0))\nonumber\\
    &-\cD_B(\rho_1)Q_1\Big)+O_\lambda(g^3) 
    \end{align}
    where $O_\lambda$ means the error term depends on $\lambda$.
\end{prop}
\begin{rem}\label{dissrho1}
    There are different ways to express some of the operators above, since the identity $\cL_g(\rho^+_g)=0$ implies
    \begin{align}
        i[H,\rho_0]=\cD_A(\rho_1)+\cD_B(\rho_1).
    \end{align}
\end{rem}
\proof:
We first note that the action of the individual dissipators on expansion (\ref{exprho+}) is of order $g$, thanks to (\ref{dissip}) and (\ref{rhozero}): for $\#\in\{A,B\}$,
\begin{align}
    \cD_\#(\rho_g^+)&=\cD_\#(\rho_0)+g\cD_\#(\rho_1)+g^2\cD_\#(\rho_2)+O(g^3)\nonumber\\
    &=g\cD_\#(\rho_1)+g^2\cD_\#(\rho_2)+O(g^3).
\end{align}

Focusing on $\sigma_A(\rho^+_g)$ only, the arguments for $\sigma_B(\rho^+)$ are similar, we deduce from expansions (\ref{exprho+}), (\ref{exprhosharp}), (\ref{explog+}) and (\ref{explogsharp}) that for any $\lambda\in \R$,
\begin{align}
    0\leq \sigma_A(\rho^+)=&g\, \tr \big((-i\lambda[H,\rho_0]+\cD_A(\rho_1))(\ln (\rho_0^A)-\ln(\rho_0)) \big)\\
    &+g^2\, \tr \Big((-i\lambda[H,\rho_0]+\cD_A(\rho_1))(\lambda Q_1^A-Q_1)\nonumber\\
    &\phantom{xxxxxxx}+(-i\lambda[H,\rho_1]+\cD_A(\rho_2))(\ln (\rho_0^A)-\ln(\rho_0)))\Big)+O_\lambda(g^3),\nonumber
\end{align}
Since $\sigma_A(\rho^+_g)\geq 0$ and the zeroth order term vanishes, the term of order $g$ needs to vanish as well, so that, reorganizing the powers in $\lambda$, 
\begin{align}
    0\leq \sigma_A(\rho^+)=
    &+g^2\, \tr \Big(-\lambda^2 i[H,\rho_0]Q_1^A\nonumber\\
    &+\lambda(\cD_A(\rho_1)Q_1^A+i[H,\rho_0]Q_1-i[H,\rho_1](\ln (\rho_0^A)-\ln(\rho_0)))\nonumber\\
    &+\cD_A(\rho_2)(\ln (\rho_0^A)-\ln(\rho_0))-\cD_A(\rho_1)Q_1\Big)+O_\lambda(g^3).
\end{align}
Then we note that, since for any $\mu^{CB}>0$, 
\be
\ln (\tau_A\otimes \mu^{CB})=\ln(\tau_A)\otimes \un_{CB}+\un_A\otimes \ln(\mu^{CB}),
\ee
we have with $X^{CB}=\ln(\rho_0^{CB})-\ln(\rho_0^B\otimes\tau_B)$,
\begin{align}
    \cD_A(\rho_2)(\ln (\rho_0^A)-\ln(\rho_0))=\cD_A(\rho_2)(\un_A\otimes
    X^{CB}).
\end{align}
Now, for any $\rho$ and $X^{CB}$,
\begin{align}\label{proptraD}
\tr \big( \cD_A(\rho)\un_A\otimes X^{CB}\big)&=
\gamma_A\tr_{CB}\tr_A\big((\tau_A\otimes \tr_A(\rho)-\rho)\un_A\otimes X^{CB}\big)\nonumber\\
&=\gamma_A\tr_{CB}(\tr_A(\rho)-\tr_A(\rho))X^{CB})=0.
\end{align}
Similar considerations yield the expression provided for $\sigma_B(\rho^+)$.
\ep
\medskip

It follows from these expansions that for any interval $I\subset \R$, there exists $g_0(I)$, such that for all $0<g<g_0(I)$, and all $\lambda\in I\setminus \{0,1\}$,
\begin{align} 
    0\leq \sigma(\rho^+_g)&=\sigma_A(\rho^+_g)+\sigma_B(\rho^+_g)\\ 
    &=g^2\big(\sigma_A^{(2)}(\lambda)+\sigma_A^{(2)}(\lambda)\big)+ O_\lambda (g^3)\nonumber\\ 
    &=g^2 \sigma^{(2)}(\lambda) + O_\lambda (g^3)
    \nonumber \\
    &=g^2\big((a_A\lambda^2+\lambda b_A+ c_A)+(a_B(1-\lambda )^2+(1-\lambda) b_B+  c_B)\big)+ O_\lambda (g^3),\nonumber
\end{align}
where superscripts $(2)$ mean second order in $g$ and subscripts $A, B$ label the reservoirs, and the coefficients $a_\#, b_\#, c_\#$ are identified by Proposition \ref{prop:secordr}.
\medskip

Thus, for $g_0(I)$ small enough, the strict positivity of $\sigma(\rho^+_g)$ is a consequence of the positivity of the second order expressions above. 
In particular, focusing on the reservoir $A$, $\sigma_A^{(2)}(\lambda)=(a_A \lambda ^2 +\lambda b_A+ c_A)\geq 0$ for all $\lambda\in I\setminus \{0,1\}$, if $g\leq g_0(I)$. Hence, unless $\sigma_A^{(2)}(\lambda)\equiv 0$ on $I\setminus \{0,1\}$, which is equivalent to $a_A=b_A=c_A=0$, $\sigma_A^{(2)}$ may vanishes at one point $\lambda_0$ at most in $I\setminus \{0,1\}$. More precisely, considering  $I$ large enough and $0\in I$, we have for $\lambda \not\in \{0,1\}$,
\begin{align}\label{difcase}
& o) \ a_A\geq 0, \ c_A\geq 0. \nonumber \\
& i) \ a_A=c_A=0 \ \Rightarrow \ b_A=0 \ \ \mbox{and} \ \ \sigma_A^{(2)}\equiv 0, \nonumber \\
    & ii) \ a_Ac_A=0 \ \Rightarrow \ b_A=0 \ \ \mbox{and} \ \ \sigma_A^{(2)}(\lambda)=
    \left\{
    \begin{matrix}
        \lambda^2a_A >0 & \mbox{if} \ c_A=0 \\ c_A >0 & \mbox{if} \  a_A=0, \, 
    \end{matrix}
    \right. 
    \nonumber\\
    & iii) \ a_Ac_A >0 \ \Rightarrow \ \ \sigma_A^{(2)}(\lambda)=\lambda ^2 a_A+\lambda b_A+ c_A
    \left\{
    \begin{matrix}
        >0 & \mbox{if} \ 4a_Ac_A>b_A^2 \\ \geq 0 & \mbox{if} \ 4a_Ac_A=b_A^2
    \end{matrix}
    \right.
\end{align}
with $\sigma_A^{(2)}$ vanishing at $\lambda_0=-b_A/2a_A\neq 0$ only, in the last case. \medskip

A similar reasoning holds for $\sigma_B^{(2)}(\lambda)$, for $I\ni 1$, large enough, where the only possible vanishing point is $1-\lambda_1=-b_B/2a_B\neq 0$, unless $\sigma^{(2)}_B\equiv 0$.
Thus, provided $\sigma_\#^{(2)}\not\equiv 0$ for $\#\in \{A,B\}$, 
$\sigma^{(2)}(\lambda)$ can only vanish for $\lambda\not\in\{0,1\}$ if  $4a_Ac_A=b_A^2>0$, $4a_Bc_B=b_B^2>0$ and $\lambda_0=\lambda_1$. {\it i.e.}
\be
2a_Aa_B+a_Ab_B+a_Bb_A=a_A(a_B+b_B)+a_B(a_A+b_A)=0.
\ee
Since $a_\#>0$, $b_A$ and $b_B$ cannot be both positive. We introduce $\varepsilon_\# =\sign(b_\#)\in\{+1, -1\}$, $\#\in\{A,B\}$ such that $b_\#=\varepsilon_\#2\sqrt{a_\#c_\#}$.
Therefore, provided the condition 
\begin{align}\label{condzero}
   1+ \varepsilon_A\sqrt{\frac{c_A}{a_A}}+\varepsilon_B\sqrt{\frac{c_B}{a_B}}=0
\end{align}
is satisfied, for $\varepsilon_A$, $\varepsilon_B$ not both positive, $\sigma^{(2)}(\lambda)=(a_A+a_B)(\lambda+\varepsilon_A\sqrt{c_A/a_A})^2$ vanishes at the point $\lambda_0=-\varepsilon_A\sqrt{c_A/a_A}\not\in \{0,1\}$.

In case $\sigma_B\equiv 0$ but $\sigma^{(2)}\not \equiv 0$, we have  $\sigma^{(2)}(\lambda)>0$, for all $\lambda\not\in\{0,1\}$, unless the second condition of iii) above holds, in which case $\sigma^{(2)}(\lambda)>0$ for all $\lambda\not\in\{0,1, \lambda_0\}$ and  $\sigma^{(2)}(\lambda_0)=0$. The similar statement holds for $A$ and $B$ exchanged with $\lambda_0$ replaced by $\lambda_1$.
\medskip

Summarizing the argument above we have finished the proof of Theorem \ref{thmmain}.
\phantom{x}\hfill \ep

\medskip

\proof of Proposition \ref{prop:critpot}:
    Actually, we prove that $c_A+c_B=0$ if and only if $[H,\rho_0]= 0$, which together with $[H,\rho_0]=0 \ \Rightarrow \ a_A=a_B=0$ yields the result by (\ref{difcase}), point i).

We first note that Proposition \ref{prop:secordr} and Remark \ref{dissrho1} yield
\begin{align}
    c_A+c_B=-\tr\big( (\cD_A(\rho_1)+\cD_B(\rho_1))Q_1\big)=-i\tr \big([H,\rho_0]Q_1\big).
\end{align}
Then we observe that
    \be \label{exlog}
    \tr \big([H,\rho_0]Q_1\big)= \big([\ln \rho_0,\rho_1] H\big).
    \ee
Indeed, using (\ref{deflogR}) we have 
\begin{align}
    [\rho_0,Q_1]&=\frac{1}{2 i \pi}\int_C [\rho_0-z, (\rho_0-z)^{-1}\rho_1 (\rho_0-z)^{-1}]\ln(z) dz\nonumber\\
    &=-[\rho_1, \ln(\rho_0)],
\end{align}
and we conclude by the identity $\tr ([A,B]C)=\tr([B,C]A)=\tr ([C,A]B)$.

To compute $[\ln \rho_0,\rho_1]$, let us recall expressions (\ref{rhozero})
\begin{align}\label{rhozerobis}
\rho_0&=\tau_A\otimes \rho_0^C\otimes \tau_B\in \mathcal{DM(H)}, \ \mbox{where} \ \rho_0^C=\diag_\tau \rho_0^C \in \ker \Phi_D\cap \mathcal{DM(H)}, \nonumber \\ 
\rho_1&=R_1+\tau_A\otimes r_C^{(1)}\otimes \tau_B, \ \mbox{with } \ R_1=i\cL_0^{-1}([H, \rho_{0}]).
\end{align}

We have
\begin{lem}
    \be \tr \big( [\ln(\rho_0), \tau_A\otimes r_C\otimes \tau_B] H \big) = 0, \ \ \forall\,  r_C \in \cB(\cH_C).\ee
\end{lem}
\proof:
Consider
\be\label{logprod}
\ln(\rho_0)=\un_A\otimes \ln(\rho_0^{C})\otimes \un_B+\ln(\tau_A)\otimes \un_C\otimes \un_B+\un_A\otimes \un_C\otimes \ln(\tau_B),
\ee
where $\ln(\rho_0^C)=\diag_\tau \ln(\rho_0^{C})$, since $\rho_0^C\in \diag_\tau \cB(\cH_C)$. Thus,
decomposing $r_C$ as 
$r_C=\diag_\tau r_C+ \offdiag_\tau r_C$,
we have 
\be
[\ln (\rho_0),\tau_A\otimes \diag_\tau r_C \otimes \tau_B]=\tau_A\otimes [\ln (\rho_0),\diag_\tau r_C] \otimes \tau_B]=0.
\ee
For the offdiagonal part, taking into account definition (\ref{hbar}), we compute
\begin{align}
    &\tr\big( [\ln (\rho_0),\tau_A\otimes \offdiag_\tau r_C \otimes \tau_B] H \big)
    =\tr\big( \tau_A\otimes [\ln (\rho_0^{C}),\ \offdiag_\tau r_C] \otimes \tau_B \, H\big)
    \nonumber\\
    &=\tr_C\big([\ln(\rho_0^C),  \offdiag_\tau r_C]\overline{H}^{\tau}]\big)=\tr_C\big([ \overline{H}^{\tau}, \ln(\rho_0^C)]\offdiag_\tau r_C \big)=0,
\end{align}
since $\rho_0^C=\diag_\tau \rho_0^C$.
\ep
\medskip

At this point, using the above, (\ref{rhozero}) and (\ref{dissinv}) we are left with
\begin{align}\label{baseq}
-i\tr \big([H,\rho_0]Q_1\big)&=\tr \big([\ln (\rho_0),\cL_0^{-1}([H,\rho_0])]H \big)\nonumber\\
&=\frac{1}{\gamma_A+\gamma_B}\Big(\tr \big(\big[\big\{ [H, \rho_0] + \frac{\gamma_A}{\gamma_B}\tau_A\otimes \tr_A ([H,\rho_0])\nonumber\\
&\phantom{\hspace{2cm}}+\frac{\gamma_B}{\gamma_A}\tr_B ([H,\rho_0])\otimes \tau_B\big\}, \ln(\rho_0) \big]H\big)
\Big).
\end{align}
Up to positive numerical factors, the first contribution of the RHS writes $\tr \big(\big[[H, K], f(K)\Big]H\big)$, for $K=\rho_0$ and $f=\ln$. As the next lemma shows, the other two contributions also have that structure. We spell out the result for the second term of the RHS only, but the analogous statement holds for the third term as well. 
\begin{lem}\label{structpart}
    With $\overline{H}^{\tau_A}=\tr_A(H\, \tau_A\otimes \un_C\otimes \un_B)$, we have
    \begin{align}
        \tr \big(\big[\tau_A\otimes \tr_A ([H,\rho_0]),\ln(\rho_0)\big]H\big)=\tr_{CB}\big(\big[[\overline{H}^{\tau_A},\tr_A( \rho_0)]\ln(\tr_A(\rho_0))\big]\overline{H}^{\tau_A}\big).
    \end{align}
\end{lem}
\proof:
We start from
\begin{align}
    \tr_A ([H,\rho_0])=\tr_A \big( [H,\tau_A\otimes \rho_0^C \otimes \tau_B]\big)= [\overline{H}^{\tau_A}, \tr_A(\rho_0)].
\end{align}
Writing (\ref{logprod}) under the form $\ln(\rho_0)=\ln(\tau_A)\otimes \un_C\otimes \un_B+\un_A\otimes \ln(\tr_A (\rho_0))$ we derive
\begin{align}
    \big[\tau_A\otimes [\overline{H}^{\tau_A}, \tr_A(\rho_0)] , \ln(\rho_0)\big]
    =\tau_A\otimes \big[[\overline{H}^{\tau_A}, \tr_A(\rho_0)], \ln(\tr_A (\rho_0))\big]
\end{align}
and 
\begin{align}
    \tr \big(\tau_A\otimes &\big[[\overline{H}^{\tau_A}, \tr_A(\rho_0)], \ln(\tr_A (\rho_0))\big]  H\big)\nonumber\\
    &=\tr_{CB}\big( \big[[\overline{H}^{\tau_A}, \tr_A(\rho_0)], \ln(\tr_A (\rho_0))\big]\overline{H}^{\tau_A}\big).
\end{align}
\ep

For the last step of the proof, we show
\begin{lem}\label{lem:tec}. Let $\cH$ be a finite dimensional Hilbert space, $H\in\cB(\cH)$ and $K\in \cB(\cH)$ be normal with spectral decomposition $K~=~\sum_{i\in \mathcal{I}}|\ffi_i\ket\bra \ffi_i | \lambda_i$, where  $\lambda_i\in \C$ are the (repeated) eigenvalues and
$\{\ffi_i\}_{ i\in\mathcal{I} }$ forms an orthonormal basis of eigenvectors. 
Let $f:\spec (K)\ra \C$ and 
$f(K)=\sum_{i\in \mathcal{I}}|\ffi_i\ket\bra \ffi_i | f(\lambda_i)$. Then
    \be 
    \tr\big(\big[[H,K], f(K)\big] H\big)=\sum_{j,k\in \mathcal{I}}\bra \ffi_j | H\ffi_k\ket\bra \ffi_k | H\ffi_j\ket (\lambda_j-\lambda_k)(f(\lambda_j)-f(\lambda_k)).
    \ee
\end{lem}
\begin{rem}
As a Corollary, 
\begin{align}\label{poscom}
H=H^*, \ K>0 \ \Rightarrow \ \tr \big(\big[[H,K], \ln(K)\big]H\big)\geq 0, \ \mbox{with equality iff} \ [H,K]=0.
\end{align}
The argument for the non-trivial implication uses 
$$\bra \ffi_j | H\ffi_k\ket\bra \ffi_k | H\ffi_j\ket= |\bra\ffi_j | H\ffi_k\ket|^2,$$ 
the fact that $\ln$ is strictly increasing, and that if $P_i$ denotes the spectral projectors of $K$, then the trace in (\ref{poscom}) being zero implies $P_i H P_j=0$ for $i\neq j$.
\end{rem}
\proof of Lemma \ref{lem:tec}:
We compute thanks to the cyclicity of the trace, and by exchanging the summation indices
\begin{align}
    \tr\big(\big[H,K], f(K)\big] H\big)&=2 \tr ( Kf(K)H^2 - f(K)HKH)\\
    &\phantom{\hspace{-3cm}}=2\sum_{j,k\in \mathcal{I}}\bra \ffi_j | H\ffi_k\ket\bra \ffi_k | H\ffi_j\ket(\lambda_jf(\lambda_j)-f(\lambda_j)\lambda_k)\nonumber\\
    &\phantom{\hspace{-3cm}}=\sum_{j,k\in \mathcal{I}}\bra \ffi_j | H\ffi_k\ket\bra \ffi_k | H\ffi_j\ket \big( (\lambda_jf(\lambda_j)-f(\lambda_j)\lambda_k)+
    (\lambda_kf(\lambda_k)-f(\lambda_k)\lambda_j)\big).\nonumber \ep
\end{align}
Coming back to (\ref{baseq}), we observe that all numerical pre-factors are  positive and all operators involved are self-adjoint. Moreover, thanks to Lemma \ref{structpart}, and the fact that $\rho_0>0$ implies $\tr_A (\rho_0)>0$ and $\tr_B(\rho_0)>0$, all contributions have the form described in (\ref{poscom}).
Moreover, $[H,\rho_0]=0\Rightarrow [\overline{H}^{\tau_\#}, \tr_\#(\rho_0)]=0$, for $\#\in\{A,B\}$, thanks to the tensorial structure (\ref{rhozerobis}) of $\rho_0$ and the definitions of $\overline{H}^{\tau_\#}$, 
so that we get $\sigma^{(2)}\equiv 0$ if and only if $c_A+c_B=0$, which ends the proof of Proposition \ref{prop:critpot}.
\ep

\section{Example on $\C^2\otimes \C^2\otimes\C^2$}\label{sec:ex3qubits}

In this section, we illustrate some of the theorems and lemmas discussed and proved above with a realistic model of an open quantum system driven by quantum resets and made of three interacting qubits in a chain. This model is particularly attractive as it is can be implemented within several solid-state experimental platforms, superconducting, semiconducting and trapped ions to cite the most relevant ones in the context of information processing. {In this work, we extend the general solution obtained within a perturbative approach in Ref.~\cite{HJ} to a more general Hamiltonian coupling of the qubits, and we investigate in particular the behaviour of entropy production in this boundary-driven open quantum system.} \\

\subsection{Model}

The model consists of a chain of three qubits $A-C-B$ characterized by their bare Hamiltonian $H_0$ with individual energies $e_A, e_C, e_B$ and interacting through $H_I$:
\bea
H_0 &=& e_A |1\ket\bra 1 | \otimes \un_C \otimes \un_B + \un_A \otimes e_C |1\ket\bra 1 | \otimes \un_B + \un_A \otimes \un_C \otimes e_B |1\ket\bra 1 |\,, \\
H_I &=& U\, (|11\ket_{AC} \bra 11 | \otimes \un_B + \un_A \otimes (|11\ket_{CB} \bra 11 | ) \nonumber \\
&& +  J_\alpha \big( |01\ket_{AC} \bra 10 | + |10\ket_{AC} \bra 01 | \big) \otimes \un_B \nonumber \\
&& + \un_A \otimes J_\beta \big( |01\ket_{CB} \bra 10 | + |10\ket_{CB} \bra 01 | \big) \,.
\eea
Here, $U, J_\alpha, J_\alpha$ are real parameters, {setting the strength of onsite and nearest-neighbour interactions respectively}. The total Hamiltonian of the three qubits is 
\begin{eqnarray}
   H = H_0 + H_I.
\end{eqnarray}

\noindent We consider the two ends ($A$ and $B$) to be subject to a reset process, with a diagonal reset state in the canonical basis of each qubit $\{ \vert 0 \rangle_{\#}, \vert 1 \rangle_{\#} \}$, $\# = A,B$:
\be
\tau_\#= 
\left( \begin{array}{cc}
t_\#&0 \\ 0 & 1-t_\#
\end{array}\right), \ \ \#\in \{A, B\}.
\ee
The dynamics which we consider follows the definitions in previous sections, \textit{i.e.} a \qrm where the unitary part is proportional to a constant $g$. Following the notation introduced in \eqref{couqrm}, the corresponding evolution equation reads: 
\begin{eqnarray}
    \dot{\rho} &=& g \cL_1(\rho) + \cL_0 (\rho) \\
&=& -i g [H, \rho] + \gamma_A (\tau_A \otimes \tr_A(\rho) - \rho) + \gamma_B (\tr_B(\rho) \otimes \tau_B - \rho)\,.
\end{eqnarray}
According to the splitting introduced in \eqref{couqrm} and \eqref{decomp3}, the evolution equation also reads: 
\begin{eqnarray}
\label{partial_ex}\nonumber
    \dot{\rho} &=&  -i g  [\lambda H, \rho] + \gamma_A (\tau_A \otimes \tr_A(\rho) - \rho) -i  g [(1-\lambda)H, \rho]\gamma_B +(\tr_B(\rho) \otimes \tau_B - \rho) \\
&=&  \cL^A_{\lambda g}(\rho) + \cL^B_{(1-\lambda)g} (\rho)\,.
\end{eqnarray}

 In the following, we will express all results in the canonical basis of the three qubits,
\begin{equation}\label{basisorder}
\{ \vert 000 \rangle, \vert 001 \rangle, \vert 010 \rangle, \vert 011 \rangle,\vert 100 \rangle,\vert 101 \rangle,\vert 110 \rangle,\vert 111 \rangle \}\,,
\end{equation}
obtained from the tensor product of the canonical basis for each qubit
\begin{equation}
    \{ \vert 0 \rangle_A , \vert 1 \rangle_A \} \otimes \{ \vert 0 \rangle_C , \vert 1 \rangle_C \} \otimes \{ \vert 0 \rangle_B , \vert 1 \rangle_B \}\,.
\end{equation}
In this basis, the two reset states are diagonal. This choice allows us to get compact and insightful expressions as discussed in the following. In particular, we observe that since $H$ commutes with the tensor product of three identical diagonal matrices in the qubit canonical basis, $\{ \vert 0 \rangle_{\#}, \vert 1 \rangle_{\#} \}$,
we get exact invariant states for the partial Lindbladians in \eqref{partial_ex} for any $g$, any $\#\in\{A,B\}$
\begin{align}
\label{sol_partial}
    \cL^\#_{g}(\tau_\#\otimes\tau_\#\otimes \tau_\#)=0.
\end{align}
This simplifies the results on entropy production as discussed below after verification of the generic assumptions.

\subsection{Generic assumptions}

We first check the generic assumptions {\bf Spec ($\overline{H}$)}, {\bf Spec ($\overline{H}^{\, \tau_\#}$)}, {\bf Coup}, {\bf Coup$^\#$}, and {\bf Pos.} for our model. The Hamiltonian $\bar{H}^\tau$ in \eqref{hbar} can be computed explicitly. In the reduced Hilbert space of qubit $C$  spanned by $\{ \vert 0 \rangle_{C}, \vert 1 \rangle_{C} \}$, it takes the form
\begin{eqnarray}
    \bar{H}^\tau = \left( \begin{array}{cc}
 e_A (1-t_A) + e_B (1- t_B) & 0 \\
 0 & (e_A + U) (1-t_A) + (e_B +U) (1-t_B) + e_C  \end{array}  \right)\,,
\end{eqnarray}
Its spectrum is simple if  $U((1-t_A)+(1-t_B))+e_C \neq 0$. Similarly, in the basis 
$$\{ \vert 00 \rangle_{CB}, \vert 01 \rangle_{CB}, \vert 10 \rangle_{CB}, \vert 11 \rangle_{CB}\}$$ of the reduced Hilbert spaces of qubits $C$ and $B$, $\overline{H}^{\, \tau_\#}$ with $\# = A$ takes the form,
\begin{align}
&\bar{H}^{\tau_A} =  \\ \nonumber
&\left( \begin{array}{cccc}
 e_A (1-t_A) & 0 & 0 & 0 \\
 0 & e_A (1-t_A) + e_B & J_\beta & 0 \\
 0 & J_\beta & (e_A + U) (1- t_A) + e_C & 0 \\
 0 & 0 & 0 & {(e_A + U) (1-t_A) + e_B + e_C + U} \end{array}  \right)\,,
\end{align}
with an identical matrix for $\bar{H}^{\tau_B}$ with indices $A$ and $B$ exchanged.
The spectrum of $\bar{H}^{\tau_A}$ is 
\begin{align}
\sigma(&\bar{H}^{\, \tau_A}) = \Big\{ e_A (1-t_A), (e_A + U)(1-t_A) + e_B + e_C + U, \nonumber\\
&(e_A +U/2)(1- t_A) + (e_B + e_C)/2 \pm \sqrt{J_\beta^2 + ( e_B -e_C - U(1-t_A))^2/4}\Big\}.
\end{align}
Importantly, the spectrum is simple and the Bohr frequencies are distinct for generic values for the parameters $\{ e_A, e_B, e_C, U, J_\beta\}\neq~0$. These results ensure the validity of {\bf Spec ($\overline{H}^{\, \tau}$)}, {\bf Spec ($\overline{H}^{\, \tau_\#}$)} for our specific model, with generic values of the parameters.\\

The assumption {\bf Coup} that the kernel of $\Phi_D(\cdot)$ is of dimension 1 was verified for the same model with $H_0=0$ in the previous work by the authors, Ref.~\cite{HJ}. The addition of $H_0$ does not change the validity of {\bf Coup}. Here, we discuss {\bf Coup$^\#$}. Remarkably, explicit expressions for \eqref{phiA} and \eqref{deffidA} can be obtained. For brevity, we do not provide them in the main text, but they are available upon request to the authors. The procedure to derive \eqref{phiA} and \eqref{deffidA} involves, in particular, defining linear maps which diagonalize {$\overline{H}^{\, \tau_A}$}, $\text{Diag}_{\tau^A}$, and $\cD_A^{-1}$ as introduced in \eqref{phiA}. After careful calculations, we get $\Phi_D^A(\cdot)$ in \eqref{deffidA}, which is a $16 \times 4$ matrix. Its kernel turns out to be of dimension 1 for generic values of the parameters  $\{ e_A, e_B, e_C, U, J_\alpha, J_\beta\}\neq~0$, so that {\bf Coup$^A$} is satisfied. A similar computation shows that  {\bf Coup$^B$} is satisfied as well. Therefore, invariant states $\rho_g^\#$ such that $\cL_g^\#(\rho_g^\#)=0$ are given by unique analytic functions of $g$, $|g|$ small. Thus \eqref{sol_partial} implies
\begin{equation}
   \rho_g^A = \tau_A \otimes \tau_A \otimes \tau_A, \ \mbox{and} \ 
   \rho_g^B = \tau_B \otimes \tau_B \otimes \tau_B, 
\ \forall g, \, |g| \ \mbox{small.}
\end{equation}
In particular, this implies that $\rho_1^\#=0$, $Q_1^\# =0$, defined in an analogous way as $Q_1$ in \eqref{exprho+}, see \eqref{explogsharp}.

With these expressions, we have verified the generic assumptions making this model valid for investigating subsequent results on entropy production behaviours.

\subsection{Steady-state from perturbation theory in the coupling parameter $g$}

The steady-state solution up to first-order in $g$ is given by
\begin{eqnarray}
    \rho_g^+ = \rho_0 + g \rho_1 +O(g^2) \,,
\end{eqnarray}
and is determined by (\ref{rhozero}). As in Ref.~\cite{HJ}, we consider no drive. However, we account for a more general situation by including onsite energies in the perturbative Hamiltonian, e.g. the dynamics accounts for a unitary evolution sets by the total Hamiltonian $g H = g(H_0 + H_I)$. The calculation of the steady-state solution follows the same steps as introduced in \cite{HJ}. The zeroth-order term $\rho_0$ with $\tr(\rho_0)=1$, reads,  
\begin{equation}\label{firsttermexample}
    \rho_0 = \tau_A \otimes \rho_C^0 \otimes \tau_B\, ,
\end{equation}
where, the expression for $\Phi_D$ yields, in the canonical basis of qubit $C$, $\rho_C^0$ given by
\begin{eqnarray}
\label{rhoC0}
    \rho_C^0 = \frac{1}{J_\alpha^2 \gamma_B + J_\beta^2 \gamma_A} \left( \begin{array}{cc}
J_\alpha^2 t_A \gamma_B + J_\beta^2 t_B \gamma_A & 0 \\
0 & J_\alpha^2 (1-t_A) \gamma_B + J_\beta^2 (1-t_B) \gamma_A
    \end{array} \right)\,.
\end{eqnarray}

We note that this expression only depends on the off-diagonal coupling terms in $H_I$, set by the strengths $J_\alpha$ and $J_\beta$. The bare energies $e_A, e_B, e_C$ and onsite interaction strength $U$ do not play a role here. The first-order term $\rho_1$, self-adjoint, satisfying $\tr(\rho_1)=0$, is also obtained after a straightforward calculation,
\begin{eqnarray} \label{secondtermexample}
 && \rho_1 = \frac{J_\alpha J_\beta (t_A - t_B)}{J_\alpha^2 \gamma_B + J_\beta^2 \gamma_A} \nonumber \\
    && \left( \begin{array}{cccccccc}
0 & 0 & 0 & 0 & 0 & 0 & 0 & 0 \\
0 & 0 & i J_\alpha t_A & 0 & 0 & 0 & 0 & 0 \\
0 & - i J_\alpha t_A & 0 & 0 & i J_\beta t_B & 0 & 0 & 0 \\
0 & 0 & 0 & 0 & 0 & i J_\beta (1-t_B) & 0 & 0 \\
0 &0 & -i  J_\beta t_B & 0 & 0 & 0 & 0 & 0 \\
0 & 0 & 0 & - i J_\beta (1-t_B)& 0 &0 & i J_\alpha (1-t_A) & 0 \\
0 & 0 & 0 & 0 & 0 & -i J_\alpha (1-t_A)& 0 & 0 \\
0 & 0 & 0 & 0 & 0 & 0 & 0 & 0    \end{array} \right)\,, \nonumber \\
&&
\end{eqnarray}

\begin{figure}
        \centering
    \includegraphics[width=0.95\textwidth]{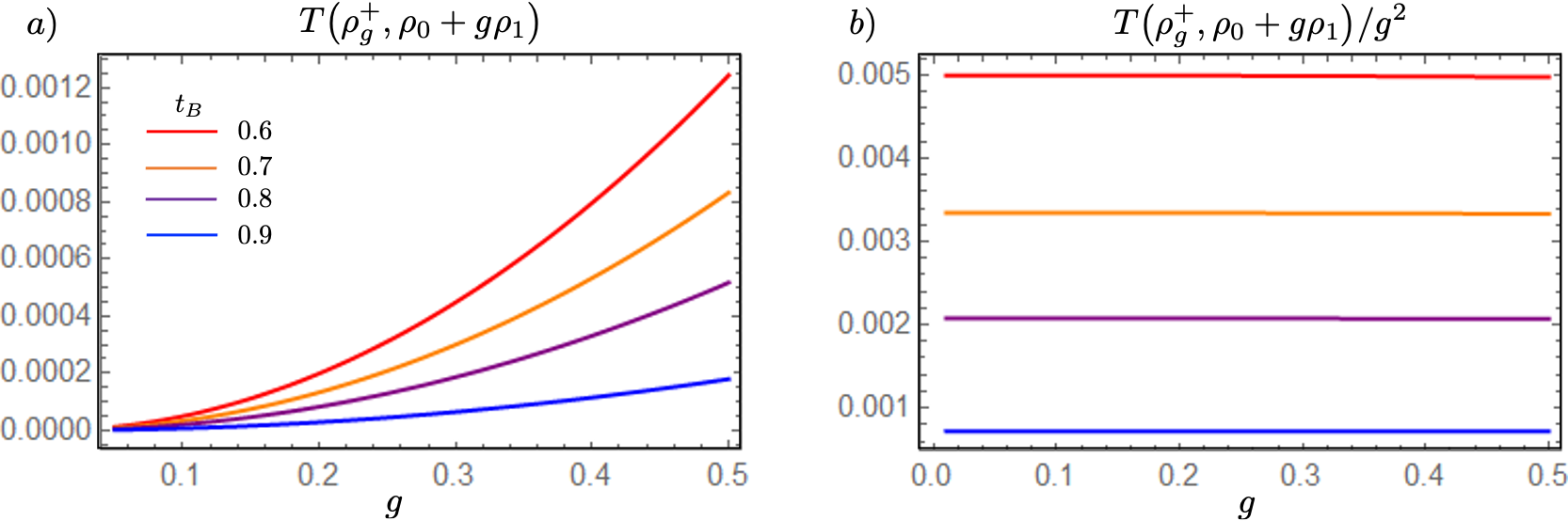}
        \caption{Trace distance $ T(\rho_g^+, \rho_0 + g \rho_1)$. a) For small $g$, the approximate solution converges to the exact solution, shown through $T(\rho_g^+, \rho_0 + g \rho_1) = O(g^2)$, as $g\rightarrow 0$, independently of the reset state characterized by $t_B = \{ 0.6, 0.7, 0.8, 0.9\}$, from top to bottom. b) Quadratic dependence on $g$ is demonstrated by plotting $T(\rho_g^+, \rho_0 + g \rho_1)/g^2$, which becomes constant as a function of $g$. Other parameters are fixed: $e_A = 0.08, e_B= 0.1, e_C = 0.05 , U=0.1, J_\alpha= 0.05 , J_\beta=0.1, \gamma_\alpha = 0.7, \gamma_\beta =0.6, t_A = 0.95$.
        }
        \label{fig:trace_distance}
    \end{figure}
    
It is interesting to remark that the first-order term vanishes when $t_A = t_B$ ($\rho_1=0$), in other words, when the two reset states are the same. This situation corresponds to $\tau_A = \tau_B = \tau$, and the zeroth-order term is then the exact solution, $\rho_0 = \tau \otimes \tau \otimes \tau$. Physically, this is an equilibrium situation. We also note that the solution up to order $O(g^2)$ does not depend on the density-density interaction strength $U$, but only on the tunneling coupling strengths between neighbouring qubits, $J_\alpha, J_\beta$, and on the reset dissipators.\\

In Fig.~\ref{fig:trace_distance}, we illustrate the validity of the expansion of the solution for small $g$ by plotting the trace distance between the exact solution $\rho_g^+$ that we obtain numerically, and the approximate solution $\rho_0 + g \rho_1$,
\begin{equation}
    T(\rho_g^+, \rho_0 + g \rho_1) = \frac{1}{2}\tr\left(\sqrt{(\rho_g^+ - (\rho_0 + g \rho_1))^\dagger  (\rho_g^+ - (\rho_0 + g \rho_1))}\right)\,.
\end{equation}
The trace distance tends to zero with $g$ as $g^2$, independently of the exact values $t_B$ characterizing the out-of-equilibrium situation due to the two \qrm. This result generalizes the perturbative result from Ref.~\cite{HJ}, as it remains valid for perturbations that include on-site energies for each qubit, $e_A, e_B, e_C \neq 0$. However, similarly to Ref.~\cite{HJ}, no drive is included beyond the perturbative approach. Below, we investigate the entropy production.

\subsection{Entropy production}

As discussed above, for this model, $\rho_1^A =0$, there is no first order corrections to $\rho_g^A$ which takes a trivial and exact form to all orders in $g$,  independent of the parameters,
   $ \rho_g^A = \tau_A \otimes \tau_A \otimes \tau_A\,.$
This result considerably simplifies the entropy productions $\sigma_A(\rho^+_g)$ in \eqref{expsigmaA+}. When $Q_1^A =0$, $\sigma_A(\rho^+_g)$ becomes an affine function of the parameter $\lambda$. To ensure its positivity $\forall \lambda$, the sum of terms proportional to $\lambda$ must be zero, and $\sigma_A(\rho^+_g)$ reduces to: 
\begin{eqnarray}
    \sigma_A(\rho^+) =  -g^2 \tr(\cD_A(\rho_1) Q_1) + O_\lambda (g^3) \,,
\end{eqnarray}
and similarly for $\sigma_B(\rho^+_g)$,
\begin{eqnarray}
    \sigma_B(\rho^+_g) = - g^2 \tr(\cD_B(\rho_1) Q_1) + O_\lambda (g^3) \,.
\end{eqnarray}

These expressions correspond to case $ii)$ of the discussion \eqref{difcase} of the order $g^2$ coefficient $\sigma_A^{(2)}(\lambda) = a_A \lambda^2 + b_A \lambda + c_A$: $Q_1^A=0$ implies $a_A=0$, which in turn imposes $b_A=0$, which yields 
\begin{eqnarray}
    \sigma_A^{(2)} = c_A >0\,, \ \text{with} \ 
   c_A = - \tr(\cD_A(\rho_1) Q_1)\,.
\end{eqnarray}

Using Remark \eqref{dissrho1}, the total entropy production, $\sigma(\rho_g^+) =  \sigma_A(\rho_g^+) +  \sigma_B(\rho_g^+)$ simplifies to
\begin{eqnarray}
\label{eq:apporx_total_EP}
    \sigma(\rho_g^+) =  -i g^2 \tr([H, \rho_0] Q_1) + O_\lambda (g^3)\,.
\end{eqnarray}

\begin{figure}
        \centering
        \includegraphics[width=0.95\textwidth]{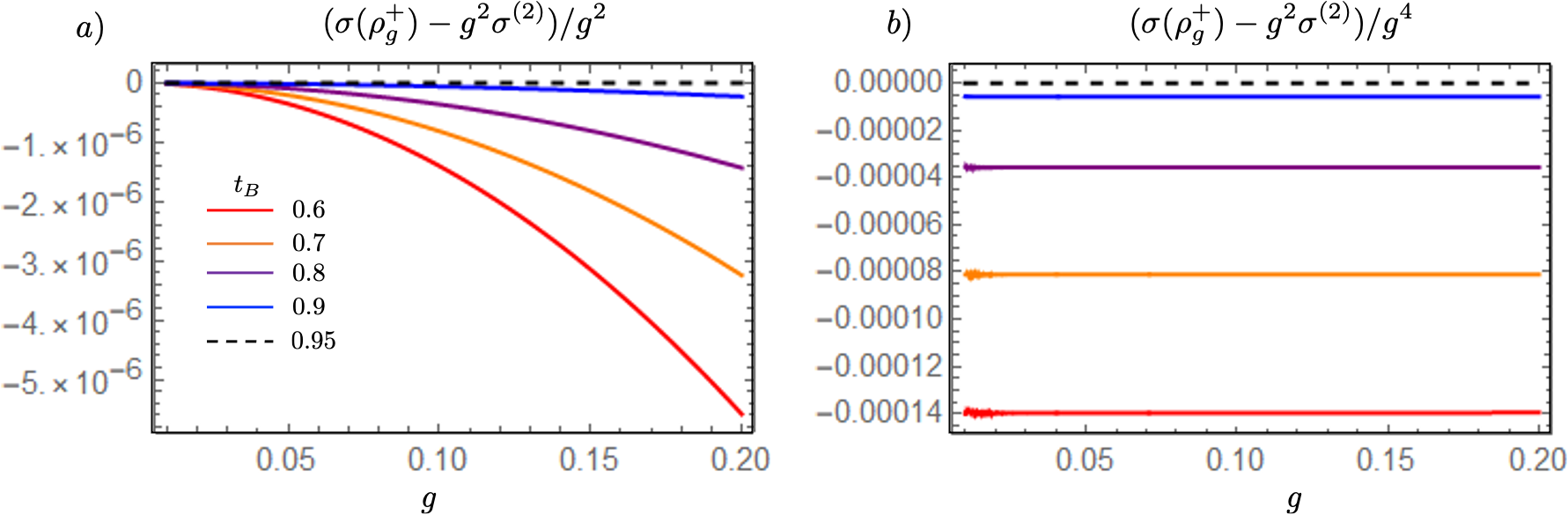}
        \caption{{\bf Theorem 7.2} for entropy production. a) We plot the difference between exact entropy production $\sigma(\rho_g^+)$ and first-order approximation $g^2\sigma^{(2)}$, see \eqref{eq:apporx_total_EP}, divided by $g^2$, as a function of $g$. We fix $t_A=0.95$, and vary $t_B$ from $0.6$ (red curve) to the equilibrium situation $t_B = 0.95$ (dashed black curve), $t_B = 0.6, 0.7, 0.8, 0.9, 0.95$. We observe that the approximate solution is larger than the exact solution, hence negative values for the difference. b) Plot of $(\sigma(\rho_g^+) - g^2\sigma^{(2)})/g^4$ to demonstrate that the next order correction term in the EP is $O(g^4)$. Other parameters are fixed: $e_A = 0.08, e_B= 0.1, e_C = 0.05 , U=0.1, J_\alpha= 0.05 , J_\beta=0.1, \gamma_\alpha = 0.7, \gamma_\beta =0.6, t_A = 0.95$.
        }
        \label{fig:thm7_2}
    \end{figure}

In Fig.~\ref{fig:thm7_2}, we illustrate {\bf Theorem 7.2} by plotting the difference $\sigma(\rho_g^+) - g^2\sigma^{(2)}$ as a function of $g$. For small $g$, this difference vanishes as $g^4$, see panel b). Whereas our approximation is shown to hold up to order $O(g^3)$ in general, Panel b) shows that for this example, at least, the error is of order $O(g^4)$. The behavior of $\sigma(\rho_g^+)$ is also a direct illustration of {\bf Proposition 7.4}, \eqref{critpot}, the entropy production is non zero \textit{iif} $[H,\rho_0] \neq 0$. For the specific model we consider, we can calculate $Q_1$ following the explicit procedure in \eqref{deflogR}, as well as  determine the conditions under which $[H, \rho_0]$ vanishes. The commutator takes the form
\begin{align}
    [H&, \rho_0] = \frac{J_\alpha J_\beta (t_A - t_B)}{J_\alpha^2 \gamma_B + J_\beta^2 \gamma_A} \times  \\
    &\left( \begin{array}{cccccccc}
    0 & 0 & 0 & 0 & 0 & 0 & 0 & 0  \\
    0 & 0 & -  J_\alpha t_A \gamma_B & 0 & 0 & 0 & 0 & 0 \\
    0 &  J_\alpha t_A \gamma_B & 0 & 0 & - J_\beta t_B \gamma_A & 0 & 0 & 0 \\
    0 & 0 & 0 & 0 & 0 & -J_\beta (1-t_B) \gamma_A& 0 & 0 \\
    0 & 0 & J_\beta t_B \gamma_A & 0 & 0 & 0 & 0 & 0 \\
    0 & 0 & 0 &  J_\beta (1-t_B) \gamma_A & 0 & 0 & - J_\alpha (1-t_A) \gamma_B & 0 \\
    0 & 0 & 0 & 0 & 0 & J_\alpha (1-t_A) \gamma_B & 0 & 0 \\
    0 & 0 & 0 & 0 & 0 & 0 & 0 & 0
    \end{array} \right) \,, \nonumber 
\end{align} 
This expression shows that, under generic assumptions, the entropy production only vanishes at equilibrium, when the two reset states are the same, $t_A = t_B$, implying for qubits $\tau_A = \tau_B$. In case of thermal baths (reset states are Gibbs states of their respective reservoir), this can be achieved when the two thermal baths are at the same temperature, $T_A = T_B \equiv T$, and qubits $A$ and $B$ are energy degenerate, $H_A = H_B = H_0$, 
\begin{equation}
    \tau_A = \tau_B = \tau = e^{-H_0/(k_B T)}/\tr(e^{-H_0/(k_B T)})\,.
\end{equation}

\begin{figure}
        \centering
        \includegraphics[width=0.95\textwidth]{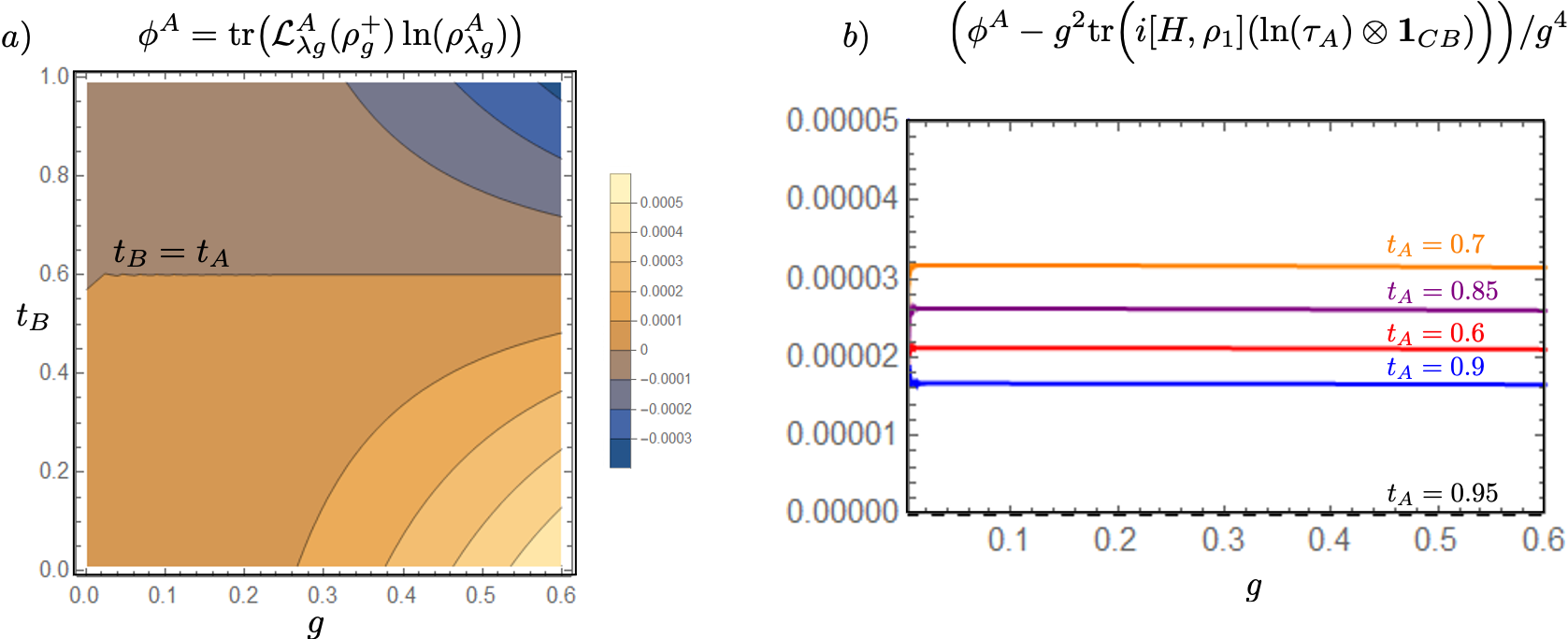}
        \caption{a) Density plot of entropy flux $\phi^A$ as a function of coupling parameter $g$ and reset ground-state population $t_B$, defined in \eqref{entropy_flux}. Entropy flux changes sign with $t_A - t_B$, appearing in the commutator $[H, \rho_1]$. b) Quartic behavior of entropy flux. The difference of the exact solution with the approximate one, divided by $g^4$, yields a constant value with $g$, but depends on the value of $t_A$. Other parameters are fixed: $e_A = 0.08, e_B= 0.1, e_C = 0.05 , U=0.1, J_\alpha= 0.05 , J_\beta=0.1, \gamma_\alpha = 0.7, \gamma_\beta =0.6$. For Panel a), $t_A = 0.6$. For Panel b). $t_B = 0.95$.
        }
        \label{fig:sigma_flux}
    \end{figure}

While the EP is always positive, the entropy fluxes are not, as shown in Fig.~\ref{fig:sigma_flux}. Similar to the total entropy, entropy fluxes vanish only for $t_B = t_A$, corresponding to an equilibrium situation with the same \qrm\ at both ends of the tri-partite system. Interestingly, these entropy currents are independent of $\lambda$, as the total EP, as shown by the following arguments. Recall the definition of the entropy fluxes
\begin{align}
    \phi^A&=\tr\big(\cL_{\lambda g}^A(\rho_g^+)\ln(\rho_{\lambda g}^A)\big), \nonumber \\
    \phi^B&=\tr\big(\cL_{(1-\lambda) g}^B(\rho_g^+)\ln(\rho_{(1-\lambda) g}^B)\big), 
\end{align}
see (\ref{entropy_flux}). We argue below that for this example, they take the simple, $\lambda-$independent forms
\begin{align}\label{simpleflux}
    \phi^A&=\tr\Big(\cD_A(\rho_g^+)(\ln(\tau_A)\otimes \un_{CB})\Big)=\gamma_A \tr \Big((\tau_A-\tr_{CB}(\rho_g^+))\ln(\tau_A)\Big), \nonumber \\
    \phi^B&=\tr\Big(\cD_B(\rho_g^+)(\un_{AC}\otimes \ln(\tau_B))\Big)=\gamma_B \tr \Big((\tau_B-\tr_{AC}(\rho_g^+))\ln(\tau_B)\Big), 
\end{align}
which, since $\sigma(\rho_g^+)=\phi^A+\phi^B$, allows us to recover that the EP is independent of $\lambda$.
Before we justify these expressions, we note that the explicit first two terms \eqref{firsttermexample} and \eqref{secondtermexample} in the expansion $\rho_g^+=\rho_0+g\rho_1+g^2\rho_2+O(g^3)$ do not contribute to the fluxes ({\it e.g.} as $\tr_{CB}(\rho_0)=\tau_A$ and $\tr_{CB}(\rho_1)=0$ and similarly for the index $B$), so that 
\begin{align}
    \phi^A&=-g^2\gamma_A \tr (\tr_{CB}(\rho_2))\ln(\tau_A))+O(g^3)=g^2\tr\Big(\cD_A(\rho_2)(\ln(\tau_A)\otimes \un_{CB})\Big)+O(g^3) \nonumber\\
    \phi^B&=-g^2\gamma_B \tr (\tr_{AC}(\rho_2))\ln(\tau_B))+O(g^3)=g^2\tr\Big(\cD_B(\rho_2)(\un_{AC}\otimes \ln(\tau_B))\Big)+O(g^3).
\end{align}
Moreover, these order $g^2$ fluxes can be expressed in terms of $\rho_1$ as
\begin{align}
\label{expgdeuxrhoun}
    \phi^A&=g^2\tr\Big(i[H,\rho_1](\ln(\tau_A)\otimes \un_{CB})\Big)+O(g^3), \ \ \mbox{and } \ \ \phi^B=g^2\tr\Big(i[H,\rho_1](\un_{AC}\otimes \ln(\tau_B))\Big)+O(g^3).
\end{align}

These expressions allow us to gain further insights on the behavior of the entropy fluxes. In particular, their signs will depend on the commutator $[H, \rho_1]$. An explicit calculation gives the expression, which depends on the bare energies $e_A, e_B, e_C$ and onsite interaction energy $U$. However, the sign is only set by the difference $(t_A-t_B)$, appearing in $\rho_1$, see \eqref{secondtermexample}. This is illustrated in Fig.~\ref{fig:sigma_flux}, where we plot the entropy flux $\phi^A$ as a function of $t_B$ for fixed $t_A$ and as function of $g$. For all $g$, $\phi^A$ changes sign according to the sign of $t_A - t_B$ and clearly vanishes at equilibrium, when $t_B = t_A$. Moreover, the behavior as a function of $g$ can be investigated beyond the results \eqref{expgdeuxrhoun}. Panel b) shows that the difference between the actual entropy fluxes and their leading order expression is of order $O(g^4)$.\\

We now justify the expressions \eqref{simpleflux}. Spelling out the Lindblad operator, and taking into account $\rho_{\lambda g}^A=\tau_A\otimes\tau_A\otimes\tau_A$, the commutator part contributes the following term to $\phi^A$ 
\begin{align}
    \tr \Big( -ig\lambda [H,\rho^+_g] \ln(\tau_A\otimes\tau_A\otimes\tau_A)\Big).
\end{align}
This term is proportional to 
\begin{align}
    \tr \Big( \rho^+_g\big[\ln(\tau_A\otimes\tau_A\otimes\tau_A),H\big]\Big), 
\end{align}
where 
\begin{align}
    \ln(\tau_A\otimes\tau_A\otimes\tau_A)=\ln(\tau_A)\otimes \un_C\otimes \un_B +\un_A\otimes \ln(\tau_A)\otimes \un_B+\un_A\otimes \un_C\otimes \ln(\tau_A).
\end{align}
This last expression commutes with $H$, and taking into account the identity (\ref{proptraD}), we get for all $g$,
\begin{align}\label{phideda}
     \phi^A&=\tr\big(\cD_A(\rho_g^+)\ln(\tau_A\otimes\tau_A\otimes\tau_A)\big)=\tr\big(\cD_A(\rho_g^+)(\ln(\tau_A)\otimes\un_C\otimes\un_B)\big),
\end{align}
is independent of $\lambda$. Similar considerations yield
\begin{align}
     \phi^B&=\tr\big(\cD_B(\rho_g^+)(\un_A\otimes\un_C\otimes\ln(\tau_B))\big).
\end{align}
Further expressing $\cD_A$, and making use of the properties of the partial trace and $\tr(\rho_g^+)=1$, we arrive at
\begin{align}
    \phi^A&=\gamma_A\tr \Big((\tau_A\otimes \tr_A(\rho_g^+)-\rho_g^+)(\ln(\tau_A)\otimes \un_C \otimes \un_B)\Big)\nonumber\\
    &=\gamma_A\Big(\tr (\tau_A \ln(\tau_A))-\tr(\tr_{CB}(\rho_g^+)\ln(\tau_A))\Big)\nonumber\\
    &=\gamma_A \tr \Big((\tau_A-\tr_{CB}(\rho_g^+))\ln(\tau_A)\Big)
\end{align}
Similarly,
\begin{align}
   \phi^B& =\gamma_B \tr \Big((\tau_B-\tr_{AC}(\rho_g^+))\ln(\tau_B)\Big),
\end{align}
which yields the statements (\ref{simpleflux}). To obtain \eqref{expgdeuxrhoun}, it is enough to consider Remark \ref{dissrho1} with indices increased by one, and identity (\ref{proptraD}), in expression \eqref{phideda}. And similarly with index $B$.\\

\noindent {\bf Acknowledgments:} GH acknowledges the Swiss National Science Foundation through the NCCR SwissMAP. AJ wishes to thank the Université de Genève for hospitality during certain stages of this work. \\

\section{Appendix}\label{erratum}

As pointed out in Remark \ref{remerr}, the conclusion of  
Proposition 6.1 in \cite{HJ} should read, with its notations,  
"Assumption {\bf Coup} holds if there exists $j\in \{ 1,\dots,n_C\}$ such that $h_{jj}(k)>0$ for all $1\leq k \neq  j \leq n_C$." \footnote{AJ wishes to thank an anonymous referee of \cite{J} for pointing out this error.}

The conclusion of Lemma 6.3, at the source of the error, needs to be corrected as follows.  "Then, $\rank {\mathfrak h} =n-1$ if $\exists 1\leq j\leq n$ such that $h_j(k)>0$, $\forall 1\leq k\neq j\leq n$" (replacing iff by if). In the proof of that lemma, (6.19), needs to be replaced by "$\det \hat {\mathfrak h}_{jj}>0$ if  $h_j(k)>0$, $\forall 1\leq k\neq j\leq n$", and (6.20) by "$\sum_{j=1}^n \det \hat {\mathfrak h}_{jj}>0$ if $\exists 1\leq j\leq n$ s.t.  $h_j(k)>0$, $\forall 1\leq k\neq j\leq n$".

\end{document}